\documentclass{aa}
\usepackage{natbib}
\usepackage{graphicx}

\newcommand{\ap}{\approx}

\newcommand{\kms}{km~s$^{-1}$}
\newcommand{\sigzero}{\ensuremath{\sigma_{0}}}
\newcommand{\ha}{H$\alpha$}
\newcommand{\hi}{\ion{H}{i}}

\newcommand{\rone}{R\ensuremath{^{\prime}_{1}}}  
\newcommand{\rtwo}{R\ensuremath{^{\prime}_{2}}}  
\newcommand{\ronetwo}{R\ensuremath{_{1}}R\ensuremath{^{\prime}_{2}}}
\newcommand{\emax}{\ensuremath{\epsilon_{\rm max}}}

\newcommand{\amax}{\ensuremath{a_{\epsilon}}}
\newcommand{\amin}{\ensuremath{a_{\rm min}}}
\newcommand{\aten}{\ensuremath{a_{10}}}
\newcommand{\lbar}{\ensuremath{L_{\rm bar}}}

\begin{document}

\title{Double-Barred Galaxies}
\subtitle{I. A Catalog of Barred Galaxies with Stellar Secondary Bars and Inner Disks}

\author{Peter Erwin}

\offprints{P. Erwin, \email{erwin@ll.iac.es}}

\institute{Instituto de Astrof\'{\i}sica de Canarias, C/ Via L\'{a}ctea s/n, 
38200 La Laguna, Tenerife, Spain}

\date{Received 27 September 2003 / Accepted 19 November 2003}

\abstract{ I present a catalog of 67 barred galaxies which contain
distinct, elliptical stellar structures inside their bars.  Fifty of
these are double-barred galaxies: a small-scale, \textit{inner} or
\textit{secondary} bar is embedded within a large-scale, \textit{outer}
or \textit{primary} bar.  I provide homogenized measurements of the
sizes, ellipticities, and orientations of both inner and outer bars,
along with global parameters for the galaxies.  The other 17 are
classified as \textit{inner-disk} galaxies, where a large-scale bar
harbors an inner elliptical structure which is aligned with the
galaxy's outer disk.  Four of the double-barred galaxies also possess
inner disks, located in between the inner and outer bars.  While the
inner-disk classification is ad-hoc -- and undoubtedly includes some
inner bars with chance alignments (five such probable cases are
identified) -- there is good evidence that inner disks form a
statistically distinct population, and that at least some are indeed
disks rather than bars.  In addition, I list 36 galaxies which
\textit{may} be double-barred, but for which current observations are
ambiguous or incomplete, and another 23 galaxies which have been
previously suggested as potentially being double-barred, but which are
probably \textit{not}.  False double-bar identifications are usually
due to features such as nuclear rings and spirals being misclassified
as bars; I provide some illustrated examples of how this can happen.

A detailed statistical analysis of the general population of
double-bar and inner-disk galaxies, as represented by this catalog,
will be presented in a companion paper.
\keywords{galaxies: structure -- galaxies: elliptical and lenticular,
cD -- galaxies: spiral -- galaxies: kinematics and dynamics}}

\maketitle

\section{Introduction}

The first hints that disk galaxies could have more than one bar
emerged in the 1970s with observations by de Vaucouleurs (1974,
1975), who identified three galaxies where the large-scale
(\textit{outer} or \textit{primary}) bar harbored a concentric,
smaller bar (the \textit{inner} or \textit{secondary} bar) in the
nuclear region.  Subsequent identifications of double-barred galaxies
include those of \citet{sandage79}, \citet{kormendy79,kormendy82a},
and \citet{schweizer80}.  Such bar-within-bar systems were generally
thought to be isolated peculiarities, and there was essentially no
theoretical interest in the topic.

Interest picked up in the late 1980s and early 1990s, spurred in part
by a new theoretical angle: the idea that nested-bar systems might
help fuel nuclear activity by efficiently driving gas into the nuclear
regions of a galaxy, or even assist in building bulges out of disk
material \citep{shlosman89,pfenniger90}.  The use of CCDs and
near-infrared imagers allowed the detection of previously unnoticed
double bars, and they began to be considered a distinct class of
galaxies worthy of investigation and modeling
\citep[e.g.,][]{bc93,friedli93,combes94}.  The latest studies, using
well-defined samples and high-resolution imaging, suggest that as many
as $\sim 1/3$ of all early-type barred galaxies may harbor secondary
bars \citep{erwin02,laine02}.  There are even galaxies which some
authors have identified as \textit{triple-barred}
\citep{w95,erwin99,laine02}, though at least some of these candidates
have turned out, on closer inspection, to be only single- or
double-barred (\nocite{erwin99}Erwin \& Sparke 1999; see
Sects.~\ref{sec:ambiguous} and \ref{sec:false} of this paper).

Inner bars are seen in both the optical
\citep[e.g.,][]{dev75,jarvis88,w95,erwin-sparke03} and the
near-infrared
\citep[e.g.,][]{shaw93,shaw95,friedli96a,mrk97,jungwiert97,greusard00,laine02}.
 This latter fact, as well as their presence in S0 galaxies devoid of
gas and dust, indicates that they are \textit{stellar} structures, and
thus at least broadly similar to ``normal,'' large-scale bars.  (Inner
\textit{gaseous} bars are sometimes seen as well, but these are not
the subject of this catalog.)

Theoretical interest now includes hydrodynamical simulations of both
observed galaxies \citep[e.g.,][]{knapen95b,ann01,eva01} and model
double-bar systems \citep{witold02,shlosman02}.  Questions concerning
the formation, dynamical stability, and evolution of double bars have
seen increasing attention from theorists
\citep[e.g.,][]{friedli93,friedli96a,davies97,witold97,
rautiainen99,witold00,rautiainen02,elzant03}.  An intriguing result
from the simulations of Rautiainen and collaborators is the suggestion
that the inner bars of double-barred systems might form
\textit{first}, in contrast to the original outside-in formation
scenario of \citet{shlosman89}.  If this is so, then inner bars would
be among the oldest dynamical structures in these galaxies, and might
provide useful clues about their formation and early history.  There
is also growing interest in spectroscopic studies specifically aimed
at the \textit{kinematics} of double-barred galaxies.  Examples
include long-slit spectroscopy by \citet{emsellem01}, CO mapping by
\citet{eva01} and \citet{petitpas02,petitpas03}, and the 2-D optical
spectroscopy of Moiseev and collaborators
\citep{moiseev02,moiseev+02}.

Thus the time seems right for a first attempt at a comprehensive
catalog of double-barred galaxies.  \citet{moiseev01} recently provided
just such a list; however, there are several ways in which it can be
improved.  The main one is that Moiseev's list is fundamentally one of
\textit{candidate} double bars, with no attempt at confirmation or
discrimination among alternate possible identifications.  This means
that some of the galaxies in the list are not, in fact, double-barred
(as indeed Moiseev 2002 concluded on the basis of 2D spectroscopy).  A
number of suggested double-bar systems in the literature -- including
some of the recent spectroscopic targets -- are either ambiguous or not
truly double-barred; so there is also a need for identifying galaxies
which can, under some circumstances, masquerade as double-barred.
Finally, we would like to know more about the general population of
double-barred galaxies: What Hubble types are they found in?  How large
and small can the inner (and outer) bars be, and what might this tell
us about how they form and evolve?  Can we identify differences between
double- and single-barred galaxies?

This paper presents a catalog of \textit{confirmed} double-bar and
inner-disk galaxies, based on detailed examinations of over a hundred
suggested candidates.  For each galaxy, I provide measurements of bar
sizes, orientations, and ellipticities in a consistent framework,
along with basic data for the host galaxies.  I also include a list of
galaxies whose double-bar status is still ambiguous or unmeasurable,
and a list of ``false'' double bars -- galaxies where nuclear rings,
spiral arms, strong dust lanes, and the like have masqueraded as
additional bars.

\section{Compilation and Sources} 

The starting point of this catalog was the survey of early-type barred
galaxies carried out by \citet{erwin02,erwin-sparke03}.  That survey,
however, found only ten double-barred galaxies, several of which had
been identified by earlier studies, and an additional eight barred
galaxies with inner disks.\footnote{See Sect.~\ref{sec:id} for the
definition of inner disks.} To expand this to a statistically more
meaningful collection, I have made a careful search for candidate
double-barred galaxies in the literature, including previous
compilations.  Among the surveys and compilations examined were those
of \citet{kormendy79}; \citet{shaw93,shaw95}; \citet{bc93};
\citet{friedli93}; \citet{w95}; \citet{friedli96a};
\citet{friedli96b}; \citet{mrk97}; \citet{jungwiert97};
\citet{marquez99}; \citet{greusard00}; \citet{martini01};
\citet{rest01}; \citet{moiseev01}; and \citet{laine02}.  Finally, a
number of unpublished observations, both ground-based and from the
\textit{HST} archive, were examined; this turned up several more
double barred galaxies.  \citep[These observations will be presented
in][in prep.]{erwin04} The goal has been to find as many plausible
candidates as possible, though it is probably guaranteed that I have
missed some previously reported candidates along the way.

Because the sources are so many and varied, and motivated by a variety
of different aims, this catalog does not represent a uniform or
complete sample.  For a variety of reasons, it is biased towards
early-type and active galaxies, and generally consists of galaxies with
larger and stronger inner bars than may be the norm, since these are
the ones most likely to be noticed and mentioned in the literature.  It
is probably not the best sample for probing, for example, the lower
limits of inner-bar sizes; on the other hand, it probably \textit{is} a
good resource for investigating the upper range of their sizes, a
question of some theoretical interest \citep{witold00,elzant03}.

\subsection{Evaluation and Winnowing}

With the candidates in hand ($\sim 125$ galaxies in all), the next
step was to determine which galaxies were actually double-barred: as I
will discuss below, there are a number of features which can
masquerade as inner (or outer) bars, particularly if ellipse fitting
is the main or only tool used to identify bars.  Because one of the
main goals of this catalog was to produce a set of homogeneous bar
measurements, I also needed galaxies where the bar characteristics had
in fact been measured, or for which publically available images
existed.  This winnowing process produced three sorts of chaff:
galaxies which could be double-barred, but for which no measurements
were available or could be made; galaxies for which the evidence for a
double bar is ambiguous (often cases where dust or lack of resolution
confuses the issue); and galaxies for which the evidence argues
\textit{against} a double bar, usually because rings, spiral arms,
and/or strong dust lanes have produced false bar signatures.  These
unconfirmed, ``ambiguous'' and ``discarded'' galaxies are listed in
Sects.~\ref{sec:ambiguous} and \ref{sec:false}, along with the reasons
for each classification.

\begin{figure}
\resizebox{\hsize}{!}{\includegraphics{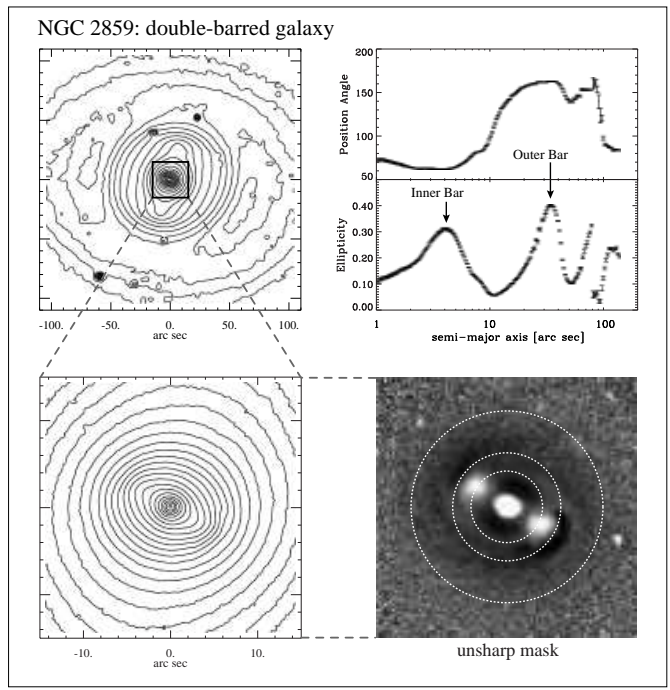}}

\caption{NGC~2859, a prototypical double-barred galaxy first noted by
\citet{kormendy79}.  The figure shows logarithmic $R$-band contours,
displaying both the outer bar and the inner bar inside it, along with
ellipse fits to the isophotes (upper right) and an unsharp mask
(lower right) of the inner bar, with hints of spiral arms just outside
it.  The dashed circles on the unsharp mask have, from the inside out,
radii = \amax, \aten, and \amin, three different ellipse-fit-based
measures of bar size; see Sect.~\ref{sec:measure} for definitions. 
(See Erwin \& Sparke 2003 for more details on the observations.)
\label{fig:n2859}}

\end{figure}

How does one verify the existence of an inner bar, or determine that
an apparent inner bar is really something else?  To start with,
Fig.~\ref{fig:n2859} gives an example of a prototypical double-bar
galaxy.  Both inner and outer bars are bisymmetric, approximately
elliptical stellar structures.  By definition, an inner or secondary
bar must appear inside a larger bar -- so part of the verification
process is ensuring that there is indeed a large-scale bar in the
galaxy.  Bars usually appear with a characterisic distortion in the
isophotes (weaker for inner bars, in part because light from the bulge
makes the isophotes rounder than they would otherwise be), usually
producing a peak in the isophotal ellipticity at the approximate
position angle (PA) of the inner bar.  A common technique for finding
bars, therefore, is to fit ellipses to the isophotes, and look for
peaks in ellipticity accompanied by relatively stationary position
angles \citep[e.g.,][]{w95,laine02}.  The problem, however, is that
there are other structures which can produce similar distortions and
similar features in ellipse fits, as I will show below \citep[see
also][]{erwin-sparke03}.  These include nuclear rings, nuclear
spirals, strong dust lanes, and star formation.  Thus, it is necessary
to go further: to inspect the images or isophotes directly, and to
make use of tools such as unsharp masking, which can help discriminate
between these different structures.

\begin{figure}
\resizebox{\hsize}{!}{\includegraphics{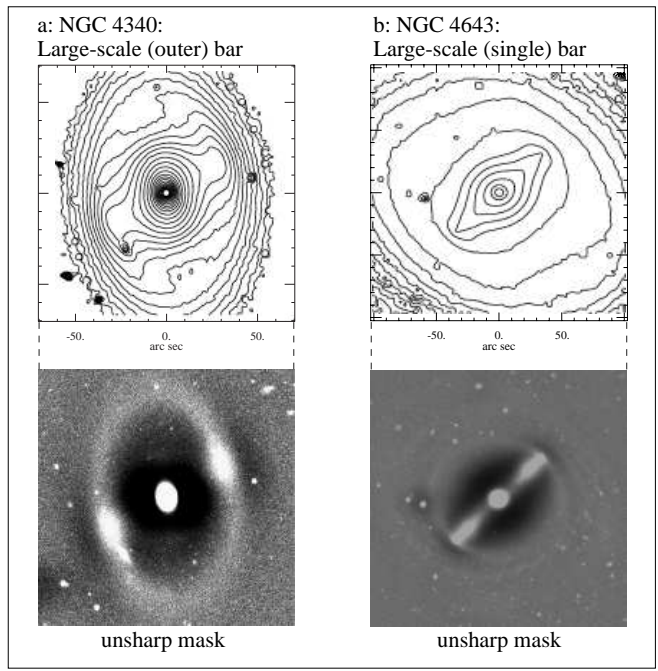}}

\caption{Examples of the appearance of ``normal'' (i.e., large-scale)
bars in unsharp masks.  For each galaxy I show logarithmically scaled
isophotes (top) and an unsharp mask from the same image (bottom). 
\textbf{a)} The outer bar of the double-barred SB0 galaxy NGC~4340;
traces of a ring surrounding the bar can also be seen ($R$-band image
from the MDM Telescope).  \textbf{b)} NGC~4643, an example of an
extremely narrow bar in an SB0/a galaxy, with the central ``spine'' of
the bar showing up clearly in the unsharp mask, along with very faint
traces of a thin ring \citep[$R$-band image from][]{erwin-sparke03}.
\label{fig:umask-bigbars}}

\end{figure}

\begin{figure}
\resizebox{\hsize}{!}{\includegraphics{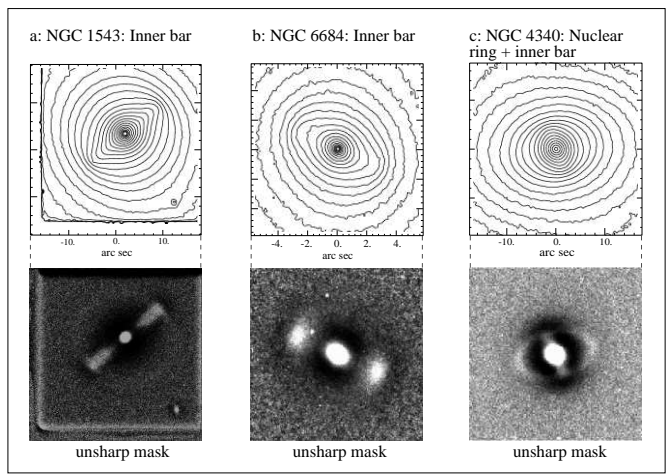}}

\caption{As for Fig.~\ref{fig:umask-bigbars}, but now showing
\textit{inner} bars from double-barred galaxies.  \textbf{a)}
NGC~1543, where the narrow inner bar is strikingly similar to the
(large-scale) bar of NGC~4643 in Fig.~\ref{fig:umask-bigbars} (WFPC2
F814W image).  \textbf{b)} NGC~6684, an inner bar with a more typical
unsharp mask appearance (WFPC2 F814W image).  \textbf{c)} NGC~4340,
where the inner bar is surrounded by an elliptical stellar nuclear
ring ($R$-band image from the MDM
Telescope).\label{fig:umask-innerbars}}

\end{figure}

As an example, Fig.~\ref{fig:umask-bigbars} shows how large-scale bars,
including the outer bars of double-barred galaxies, appear in unsharp
masks.  Depending on how narrow and strong the bar is, the unsharp mask
can show the ridgeline of the bar (e.g., NGC~4643 in the figure), or
just the characteristic sharp ends typical of so-called ``flat'' or
``early-type'' bars, where the bar's major-axis luminosity profile
steepens abruptly \citep{kormendy82b,ee85}.  What
Fig.~\ref{fig:umask-innerbars} shows is that inner bars have a very
similar appearance to large-scale bars in unsharp masks.  This is an
indication that inner and outer bars are structurally similar
\citep[][hereafter Paper~II]{erwin04-paper2}, and it also suggests that
unsharp masking can be used as a secondary technique for identifying
and confirming the presence of inner bars.

\begin{figure}
\resizebox{\hsize}{!}{\includegraphics{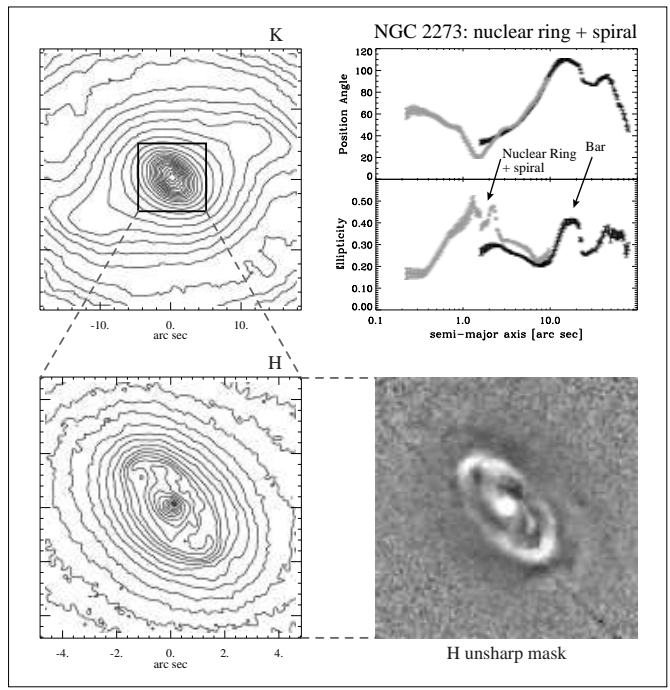}}

\caption{Example of a \textit{single}-barred galaxy where a nuclear
ring masquerades as an inner bar.  The SBa galaxy NGC~2273 was listed
as double-barred by \citet{mrk97} on the basis of ellipse fits (upper
right: black = $K$-band fits to their image, gray = \textit{HST}
$H$-band fits) and the appearance of their $K$-band image (upper
left).  However, an \textit{HST} NICMOS2 $H$-band image (lower left)
and its unsharp mask (lower right) show that the ``inner bar'' ($a
\sim 2\arcsec$) is actually a nuclear ring, with a nuclear spiral
inside; WFPC2 images and color maps confirm this
\citep[see][]{erwin-sparke03,martini03-apjs}.
\label{fig:nucrings1}}

\end{figure}

\begin{figure}
\resizebox{\hsize}{!}{\includegraphics[scale=0.9]{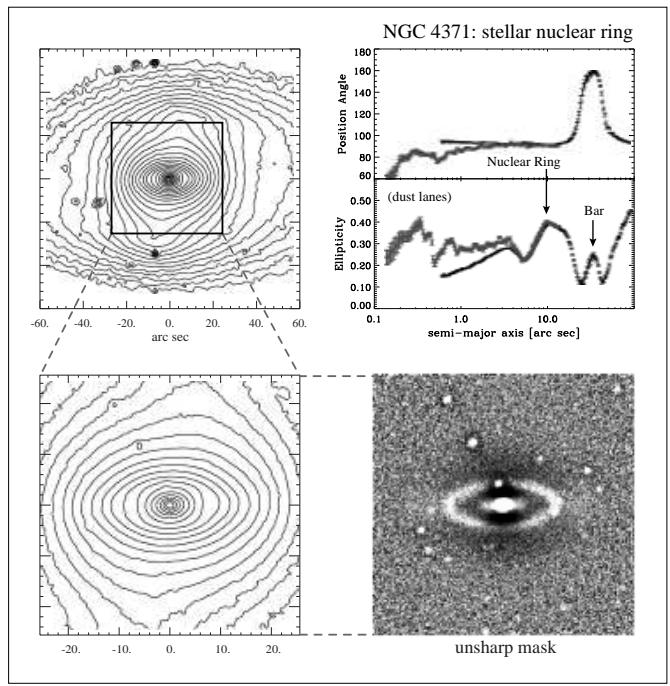}}

\caption{Another example of a nuclear ring masquerading as an inner
bar.  Here, the SB0 galaxy NGC~4371 shows three distinct peaks in
ellipse fits to ground-based images \citep[upper right: black =
ground-based $R$-band fits, gray = WFPC2 $V$-band fits; both
from][]{erwin99}, which led \citet{w95} to suggest that it was
\textit{triple}-barred.  The $R$-band images (upper and lower left) do
indeed show a prominent inner elliptical feature with $a \sim
10\arcsec$, corresponding to the middle ellipticity peak; but unsharp
masking (bottom right) shows that this is due to a stellar nuclear
ring.  The innermost ellipticity peak (in the ground-based ellipse
fits) is actually due to the combination of seeing effects and the
superposition of an elliptical ring on top of rounder bulge isophotes
\citep[see][]{erwin99,erwin01-lapalma}.  Circumnuclear dust lanes
produce variable ellipticity and position angles in the \textit{HST}
fits at $a < 1\arcsec$.
\label{fig:nucrings2}}

\end{figure}

Two examples of how nuclear rings can masquerade as inner bars are
shown in Figs.~\ref{fig:nucrings1} and \ref{fig:nucrings2}.  In both
cases, there is a clear peak in the ellipticity, with a stationary or
extremal value of the position angle, inside the large-scale
(``outer'') bar; these features are seen in near-infrared or dust-free
$R$-band images, and so are unlikely to be due to dust lanes (though it
must be kept in mind that sufficient dust can affect even $K$-band
images).  So on purely ellipse-fit criteria these are multiply barred
galaxies.  However, unsharp masking shows that the inner elliptical
features are clearly rings, with no evidence for barlike structures
inside the rings.

\begin{figure}
\resizebox{\hsize}{!}{\includegraphics{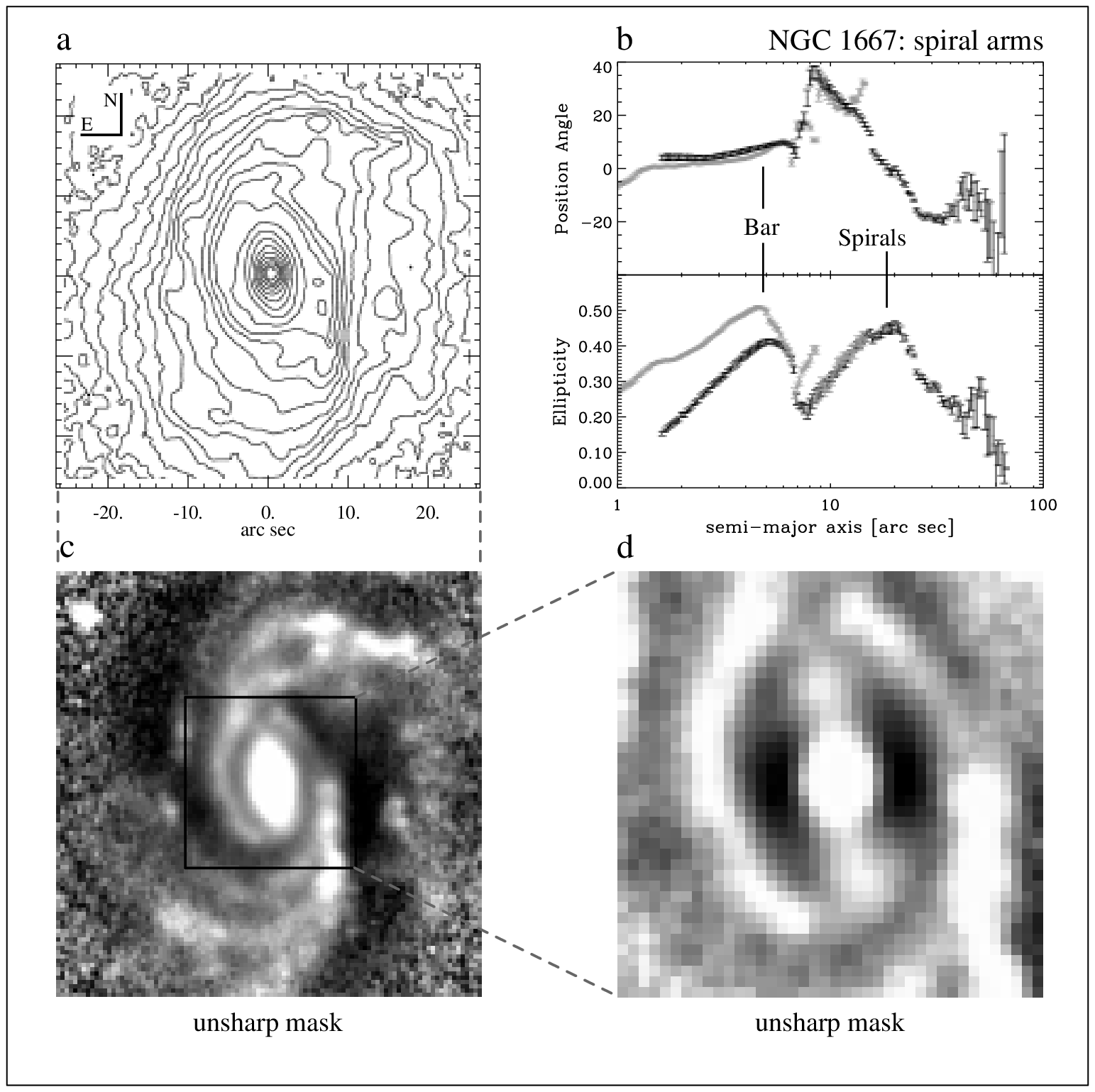}}

\caption{Another example of how ellipse fits can misleadingly suggest
two bars in a single-barred galaxy: in this case, a false ``outer
bar'' signature is produced by spiral arms in the SABc galaxy
NGC~1667.  \textbf{a)} Logarithmically spaced $K$-band isophotes
\citep[from the image of][]{mrk97}, along with ellipse fits
\textbf{(b)} to the same image (black points) plus ellipse fits to the
NICMOS2 $H$-band image (gray points).  Two strong ellipticity peaks
are present, suggesting a possible double-barred galaxy
\citep[e.g.,][]{laine02}: an ``outer bar'' with $a \sim 20\arcsec$ and
an ``inner bar'' with $a \sim 5\arcsec$.  However, unsharp masking
\textbf{(c)} with Gaussian $\sigma = 10$ pixels shows that the ``outer
bar'' feature is really due to spiral arms (Mulchaey et al.).  In
contrast, unsharp masking with $\sigma = 3$ pixels \textbf{(d)} shows
that the ``inner bar'' is indeed a bar -- almost certainly the
galaxy's \textit{only} bar.
\label{fig:n1667}}

\end{figure}

Sometimes the inner bar is real, but the \textit{outer} bar is not:
Fig.~\ref{fig:n1667} demonstrates how spiral arms can produce a false
bar signature in the ellipse fits.  In other cases, the presence of an
outer bar may be uncertain, usually due to confusion from dust and
spiral arms, or because the putative outer bar is so round or
peculiarly shaped that it isn't clear whether it is a bar or not; this
can include structures sometimes called ``oval disks''
\citep[e.g.][]{kormendy-norman79}.  In two such oval-disk galaxies
(NGC~1068 and NGC~4736), there is both good evidence for small-scale
bars \textit{and} sufficient \textit{kinematic} evidence that the oval
disks are dynamically barlike, so I consider these to be double-barred
galaxies (albeit with rather weak, atypical outer bars).

\section{Definitions and Measurements} 

\subsection{How to Measure Bars}
\label{sec:measure}

The three basic measurements one can make of a bar (or inner disk) are
its strength, its size, and its orientation.  The latter is, in
principle, the simplest and least ambiguous measurement, and is
defined as the position angle (PA), measured east from north (though
there are some caveats about \textit{how} to measure it; see below). 
The other two measurements are more difficult, both to define and to
make.  Bar ``strength'' is probably the most ambiguous and
contentious.  As \citet{athan03} notes, ``Although the notion of bar
strength is clear to everyone, and it is very often easy, when
comparing two bars, to say which one is strongest, a precise
definition is not trivial.''  I have defaulted to using one of the
simplest definitions -- the maximum ellipticity (\emax) of the
isophotes in the bar region -- because it is used rather often for
bars \citep[e.g.,][]{m95,w95,jungwiert97,laine02,erwin-sparke03} and
because it is easy to determine from published ellipse fits; this lets
me determine bar strengths for as many of the galaxies as possible. 
\citet{laurikainen02} have recently shown that bar \emax{} correlates
quite well with the more sophisticated (and difficult) measurement
based on maximum relative tangential force due to the bar 
\citep{buta-block01}.

For measuring the lengths of bars, I adopted a stratagem similar to
those of \citet{erwin-sparke03} and \citet{erwin03-bigbars}, which is
to define \textit{two} radial sizes: \amax{} and \lbar.  From the
ellipse fits, I measured the semi-major axis at maximum ellipticity
(\amax); the semi-major axis at the first minimum in ellipticity
outside the bar \citep[\amin; based on the approach of ][]{w95}; and
the semi-major axis \aten{} where the PA has changed by 10\degr{} from
the bar PA. All such values, when they can be measured, are listed in
Table~\ref{tab:measured}.  The maximum-ellipticity length \amax{} is
adopted as one measure of the bar size, in part because it is a
relatively simple and repeatable measurement and is used rather often,
including for some $n$-body bars
\citep[e.g.,][]{wozniak91,jungwiert97,laine02,athan-misir02}.  It
generally indicates the radius at which the isophotal distortion due to
the bar is at a maximum; this is often close to the radius where the
bar's major-axis profile turns over, at least in early-type galaxies.

In some cases, there is no clearly defined maximum ellipticity for the
bar.  Usually, however, there \textit{is} a local maximum or minimum
in the \textit{position angle}, which again represents the point where
the bar has its strongest effect on the isophotes.  This semi-major
axis then becomes \amax.  (See \nocite{nieto92}Nieto et al.\ 1992,
\nocite{busarello96}Busarello et al.\ 1996, and Erwin \& Sparke 2003
for examples.)

Despite the common use of \amax, there is good evidence that it is
usually an \textit{underestimate} of the bar's true length
\citep[e.g.,][]{w95,laurikainen02,athan-misir02,erwin-sparke03}; see
Fig.~\ref{fig:n2859}.  So Erwin \& Sparke defined a second
measurement, which I refer to here as \lbar{}, by taking the
\textit{smaller} of \amin{} and \aten.  In certain cases, I use an
alternate measurement to define \lbar, such as the size of an inner or
nuclear ring surrounding the bar, when it is clear that the bar does
not extend past the ring and that \amin{} and \aten{} are well outside
the bar or simply cannot be defined (these cases are noted in
Sect.~\ref{sec:notes}); the presence of such a ring can distort the
ellipse fits and produce misleadingly large values of \amin{} and
\aten.

In \citet{erwin03-bigbars}, I find that \amax{} and \lbar, as defined
above, correlated extremely well for the primary (or single) bars of a
sample of S0--Sb galaxies.  The same is true here: the Pearson and
Spearman correlation coefficients are $r = 0.93$ and $r_{s} = 0.95$ for
inner and outer bars, respectively.  The mean ratio of $\amax / \lbar$
is $0.75 \pm 0.12$ for the inner bars and $0.79 \pm 0.12$ for the outer
bars.  This is almost identical to the ratio for S0--Sb primary or
single bars (0.80) and similar to the mean ratio of
\nocite{athan-misir02}Athanassoula \& Misiriotis' (2002) L(a/b) and
L(phase) bar-size measurements (0.73), applied to bars in $n$-body
simulations.

Most of the bar measurements are based, directly or indirectly, on
fitting ellipses to the isophotes.  The actual ellipse-fit values come
from three sources: ellipse fits performed by myself on available
images; examination of published ellipse fits; and reported values
when the actual fits were not published.  Sources for the latter two
are given for each galaxy in Sect.~\ref{sec:notes}.

Although bar position angle is, as noted above, generally the simplest
and least ambiguous measurement, there can be problems if ellipse fits
are being used for bar measurements.  Ellipse-fit position angles have
the advantage of being (usually) precise and repeatable; however, it
must be remembered that they are fits to isophotes, and are sometimes a
compromise.  Multiple components (bar + bulge, or bar + spirals) and
projection effects can conspire to produce isophotes for which the
best-fit ellipse has a position angle which does not match any of the
individual components.  As discussed and illustrated by
\citet{erwin-sparke03}, there are numerous cases where the position
angle of the bar, based on its orientation in the image, does not match
the position angle from the ellipse fits.  They found that the
ellipse-fit position angle for bars was off by $> 5\degr$ for about
one-third of the large-scale (outer or single) bars in their sample.
The problem is \textit{worse} for inner bars, in part because their
isophotes are usually contaminated by light from the inner bulge,
nuclear rings, and inner disks: Erwin \& Sparke found that inner-bar
position angles differed by $> 10\degr$ from the ellipse-fit position
angles \textit{for about half of their double-barred galaxies}.
Because of these problems, all bar position angles have been checked,
wherever possible, by direct inspection of the isophotes and unsharp
masks.  If this indicated that the ellipse-fit PA value was off by more
than $\sim 2$--3\degr, then the ellipse-fit value was discarded and the
bar PA was measured directly from the image.

\textit{A simple example}: Consider the case of NGC~2859
(Fig.~\ref{fig:n2859}).  The ellipse fits show two clear peaks in
ellipticity, with $\amax = 4.1\arcsec$ and $\amax = 34\arcsec$ for the
inner and outer bars, respectively (the peaks at larger radii are due
to the lens and outer ring).  The isophotes clearly indicate bars
rather than spirals, and unsharp masking supports this.  For the inner
bar, there is an unambiguous minimum in ellipticity at $\amin =
11\arcsec$; but since the position angle has changed by $10\degr$ from
the bar PA (62\degr) by the time $a = 6.2\arcsec$, I adopt the latter
value (\aten) as \lbar.  (There \textit{is} a faint, dusty nuclear
ring surrounding the inner bar -- see Erwin \& Sparke 2003 -- but
since it has $a = 7\arcsec$, which is $> \aten$, I keep $\lbar =
\aten$.)  For the outer bar, there is a clear ellipticity minimum 
at $\amin = 52\arcsec$, but again \aten{} is smaller (43\arcsec) than 
\amin, so \lbar{} is set = \aten.

\subsection{Inner Disks} 
\label{sec:id}

Following \citet{erwin02,erwin-sparke03}, I classify some of the inner
elliptical structures as ``inner disks.''  The criteria for this are:
measured ellipticity less than that of the outer disk, measured
position angle differing by $\leq 10\degr$ from that of the outer
disk, and no evidence that the structure is a nuclear ring or spiral
(e.g., from unsharp masking or color maps).  As pointed out by Erwin
\& Sparke, this makes for a heterogeneous, hodge-podge of a category,
which can include genuine disks, inner bars with chance alignments,
unresolved stellar rings, and flattened bulges.  Erwin \& Sparke
(2002) were able to find statistical evidence that the inner disks
constituted a distinct category, different from inner bars (primarily
because they are larger relative to the outer bars); Paper~II shows
that this statistical distinction is even stronger when the much
larger set of galaxies in this catalog is considered.  There is also
evidence that at least some of the inner disks really are
\textit{disks}: \citet{erwin03-id} showed that the inner disks of
NGC~2787 and NGC~3945 are morphologically, photometrically, and
kinematically disklike.

Nevertheless, some of the ``inner disks'' presented here probably
\textit{are} inner bars which happen to be aligned with the outer
disk, particularly when the combination of inner-bar and bulge light
produces composite isophotes which are rounder than the inner bar
would be by itself.  (NGC~2642 and NGC~7098, on the other hand, are
examples of double-barred galaxies where the inner bar is closely
aligned with the outer disk, but is clearly \textit{more} elliptical;
thus, they are \textit{not} classified as inner disks.)  A few
candidates can be identified on the basis of unsharp masking -- that
is, unsharp masking suggests a typical bar structure, rather than the
smoother, elliptical appearance of most inner disks; compare
Figs.~\ref{fig:umask-bigbars} and \ref{fig:umask-innerbars} with
Fig.~1 of \citet{erwin03-id}.  These galaxies are indicated by notes
in Table~\ref{tab:galaxies-id}.

\subsection{Deprojection} 

To compare the sizes and orientations of bars and disks, I deproject
their measurements using the outer-disk inclinations and positions
angles from Cols.~5 and 6 of Tables~\ref{tab:galaxies-db} and
\ref{tab:galaxies-id}, assuming that bars are relatively flat, linear
structures.  Recent near-IR studies of edge-on galaxies
\citep{lutticke00} show that large-scale bars are indeed flat over
most of their length, with only the inner $\sim 30$--40\% of the bar
being thick (in the form of peanut-shaped bulges).  Whether
\textit{inner} bars would have their own vertical thickening is
unknown, though the high pattern speeds suggested for inner bars would
probably preclude vertical Lindblad resonances outside the very
nucleus.

Galaxies with $i < 25\degr$ are not deprojected, both because
deprojection has little effect for inclinations that small (lengths
increase by 10\% at the very most) and because such galaxies often do
not have a well-defined major-axis position angle for the outer disk.

\section{The Catalog} 

The final catalog is presented in Tables~\ref{tab:galaxies-db} and
\ref{tab:galaxies-id}, divided into double-barred galaxies and
inner-disk galaxies.  Measurements for the bars (outer and inner) and
inner disks are given in Table~\ref{tab:measured}, with deprojected bar
sizes and relative position angles ($\Delta$PA) in
Table~\ref{tab:deproj}.  For $\Delta$PA, I follow \citet{bc93} and use
a convention where the inner bar (or inner disk) ``leads'' or
``trails'' the outer bar, depending on the sense of rotation given by
the spirals (Col.~8 of Tables~\ref{tab:galaxies-db} and
\ref{tab:galaxies-id}); see Fig.~5 of Buta \& Crocker.  Since NGC~2681
has \textit{three} bars, the relative position angle is harder to
define -- e.g., which bar counts as the ``outer bar'' with respect to
the innermost bar?  -- so there is no $\Delta$PA for that galaxy in
Table~\ref{tab:deproj}.  The galaxy is almost face-on ($i \approx
18\degr$), so the interested reader can determine the various possible
relative position angles using the observed bar position angles in
Table~\ref{tab:measured}, without needing to do a deprojection.

Details and special notes for these galaxies are given in
Sect.~\ref{sec:notes}.  This includes discussions of sources for
identifications and measurements, how the outer-disk orientation was
derived -- e.g., from the outer isophotes, from \hi{} kinematics,
etc.\  -- and other items of interest.

There are a number of other candidate double-barred or inner-disk
galaxies which I do not list in the tables; these are discussed in
Sect.~\ref{sec:ambiguous}.  These are primarily cases where good
measurements of one or both bars are not yet available, or where
alternate explanations -- strong dust lanes, nuclear rings, etc.\ --
cannot be ruled out using the images currently available.  The most
promising candidate (i.e., galaxies which are probably double-barred,
but for which good measurements are not available) are NGC~2442,
NGC~4274, and IC~454.  One particularly interesting, though admittedly
ambiguous, case is the Milky Way itself, which \citet{alard01} recently
suggested might be double-barred.

Finally, candidate galaxies which were rejected as being
\textit{single}-barred (or even unbarred) are listed in
Sect.~\ref{sec:false}, along with the reasons for rejection.

A statistical analysis of these measurements will be presented in 
Paper~II.

\begin{table*}
\begin{minipage}{126mm}
    \caption{General Data for Double-Barred Galaxies}
    \label{tab:galaxies-db}
    \begin{tabular}{llrrrrrrlr}
\hline
Name & Type (RC3) & $B_{t,c}$ &  $R_{25}$ & $i$     & Disk PA & Distance
     & \sigzero{} & Rot & Notes \\
     &            &           & \arcsec{} & \degr{} & \degr{} &   Mpc
     &  \kms{}    &     &       \\
(1)  & (2)        & (3)       & (4)       & (5)     & (6)     & (7)
     & (8)        & (9) & (10) \\
\hline
 NGC 357 &               SB(r)0/a & 12.49 &   72 &   37 &   20 &  31.6 & \ldots  & $-$ & \\
 NGC 718 &                SAB(s)a & 12.34 &   71 &   30 &    5 &  22.6 &  127 & + & \\
NGC 1068 &             (R)SA(rs)b &  9.23 &  212 &   40 &   98 &  14.7 &  148 & + & \\
NGC 1097 &         (\rone)SB(rl)b &  9.76 &  279 &   46 &  134 &  14.3 & \ldots  & $-$ & \\
NGC 1241 &                SB(rs)b & 12.11 &   84 &   55 &  145 &  52.1 & \ldots  & + & \\
NGC 1291 &        (\rone)SB(l)0/a &  9.49 &  293 &    6 & \ldots  &   7.9 &  186 & + & \\
NGC 1317 & (R$^{\prime}$)SAB(rl)0/a & 11.74 &   83 &   30 &   78 &  19.3 & \ldots  & $-$ & \\
NGC 1433 &        (\rone)SB(rs)ab & 10.54 &  194 &   33 &   21 &  10.8 & \ldots  & $-$ & \\
NGC 1543 &          (R)SB(l)$0^0$ & 11.39 &  147 &   20 & \ldots  &  20.0 &  158 & \ldots  & \\
NGC 1808 &        (\rone)SAB(s:)b & 10.30 &  194 &   50 &  133 &  10.2 &  148 & + & \\
NGC 2217 &       (R)SB(rs)$0^{+}$ & 11.36 &  134 &   22 &    5 &  18.9 &  232 & + & \\
NGC 2642 &                SB(r)bc & 13.07 &   61 &   24 &  140 &  56.8 & \ldots  & $-$ & \\
NGC 2646 &             SB(r)$0^0$ & 13.53 &   40 &   21 & \ldots  &  52.2 &  222 & \ldots  & \\
NGC 2681 & (R$^{\prime}$)SAB(rs)0/a & 10.98 & 109 &  18 &  140 &  17.2 &  111 & + & 1 \\
NGC 2859 &          (R)SB(r)$0^+$ & 11.71 &  128 &   25 &   90 &  24.3 &  177 & + & \\
NGC 2950 &          (R)SB(r)$0^0$ & 11.73 &   80 &   48 &  120 &  14.9 &  183 & + & \\
NGC 2962 &        (R)SAB(rs)$0^+$ & 12.72 &   79 &   53 &   10 &  30.0 &  192 & $-$ & \\
NGC 3081 &       (\rone)SAB(r)0/a & 12.71 &   63 &   34 &   97 &  29.9 & \ldots  & + & \\
NGC 3275 &                 SB(r)a & 11.98 &   85 &   42 &  122 &  40.5 & \ldots  & + & \\
NGC 3358 &        (\rtwo)SAB(l)ab & 11.85 &   99 &   42 &  125 &  37.3 & \ldots  & $-$ & \\
NGC 3368 &              SAB(rs)ab &  9.75 &  228 &   50 &  172 &  10.5 &  114 & $-$ & 2\\
NGC 3393 &  (R$^{\prime}$)SB(s)ab & 12.79 &   66 &   25 & \ldots  &  48.3 &  184 & + & \\
NGC 3941 &             SB(s)$0^0$ & 11.13 &  104 &   51 &   10 &  12.2 &  175 & +? & \\
NGC 3945 &         (R)SB(rs)$0^+$ & 11.54 &  158 &   50 &  158 &  19.3 &  165 & $-$? & 2\\
NGC 4303 &              SAB(rs)bc & 10.04 &  194 &   25 &  138 &  15.3 &   96 & $-$ & \\
NGC 4314 &                SB(rs)a & 11.28 &  125 &   25 &   65 &  12.0 &  117 & + & \\
NGC 4321 &               SAB(s)bc &  9.79 &  222 &   27 &  153 &  15.2 &   94 & + & \\
NGC 4340 &             SB(r)$0^+$ & 11.99 &  105 &   50 &   95 &  15.3 &  113 & $-$ & \\
NGC 4503 &               SB$0^-$: & 11.82 &  106 &   64 &   12 &  15.3 &  118 & + & \\
NGC 4725 &               SAB(r)ab &  9.74 &  321 &   42 &   40 &  12.4 &  155 & $-$ & \\
NGC 4736 &             (R)SA(r)ab &  8.68 &  337 &   35 &  113 &   5.2 &  136 & $-$ & \\
NGC 4785 & (R$^{\prime}$)SAB(r)ab & 11.80 &   59 &   64 &   81 &  46.4 &  186 & $-$ & 2\\
NGC 4984 &        (R)SAB(rs)$0^+$ & 11.96 &   84 &   40 &   15 &  15.6 & \ldots  & \ldots  & \\
NGC 5365 &          (L)SB(s)0$^-$ & 11.92 &   89 &   48 &    4 &  30.5 &  232 & + & \\
NGC 5728 &         (\rone)SAB(r)a & 11.79 &   93 &   40 &    2 &  36.9 &  209 & $-$ & \\
NGC 5850 &                 SB(r)b & 11.45 &  128 &   30 &  163 &  35.2 & \ldots  & + & \\
NGC 6654 & (R$^{\prime}$)SB(s)0/a & 12.63 &   79 &   44 &    0 &  28.3 & \ldots  & $-$ & \\
NGC 6684 &          (L)SB(r)$0^+$ & 11.00 &  119 &   51 &   36 &  13.9 &  110 & \ldots  & \\
NGC 6782 &        (\rone)SB(r)0/a & 12.26 &   66 &   41 &   45 &  48.5 &  140 & + & \\
NGC 7098 &             (R)SAB(r)a & 11.94 &  122 &   42 &   75 &  28.0 & \ldots  & $-$ & \\
NGC 7187 &         (R)SAB(r)$0^+$ & 13.57 &   41 &   26 &  134 &  33.8 & \ldots  & + & 2\\
NGC 7280 &            SAB(r)$0^+$ & 12.80 &   66 &   48 &   72 &  24.3 &  111 & +? & \\
NGC 7716 &               SAB(r)b: & 12.63 &   64 &   34 &   35 &  34.1 & \ldots  & + & \\
 Mrk 573 &       (R)SAB(rs)$0^+$: & 14.49 &   40 &   12 & \ldots  &  68.3 &  130 & $-$ & \\
 UGC 524 &   (R$^{\prime}$)SB(s)b & 14.39 &   31 &   21 &  120 & 145.0 & \ldots  & + & \\
ESO 215-G031 &          (\rone)SB(r)b & 12.18 &   70 &   42 &  130 &  33.6 & \ldots  & + & \\
ESO 320-G030 &         (\rone)SAB(r)a & 13.16 &   67 &   54 &  121 &  40.1 & \ldots  & $-$ & \\
ESO 443-G039 &                    S0? & 13.43 &   40 &   58 &   14 &  39.1 &  132 & \ldots  & \\
ESO 447-G030 &        (R)SAB(rl)$0^+$ & 12.98 &   43 &   50 &   35 &  37.6 &  163 & $-$ & \\
IRAS 03565+2139 &                   SBa? & 14.91 &   17 &   20 & \ldots  & 101.0 & \ldots  & $-$ & 3\\

\hline
    \end{tabular}

\medskip

Col.\ (1): Galaxy name.  Col.\ (2): Morphological type from NED
(primarily from RC3).  Col.\ (3): Total corrected $B$-band magnitude,
from LEDA. Col.\ (4): One-half of the $\mu_{B} = 25$ diameter from
RC3.  Col.\ (5): Galaxy inclination.  Col.\ (6): Position angle of
galaxy major axis.  Col.\ (7): Distance in Mpc.  Col.\ (8): Central
velocity dispersion.  Col.\ (9): Sense of spiral arm rotation (``+'' =
clockwise, `` $-$'' = counter-clockwise).  Col.\ (10): Notes for
individual galaxies: 1 = Triple-barred galaxy.  2 = Also has an inner
disk.  3 = Morphological type is visual estimate (none listed in NED). 
Sources of data for individual galaxies are discussed in
Sect.~\ref{sec:notes}.
\end{minipage}
\end{table*}

\begin{table*}
\begin{minipage}{126mm}
    \caption{General Data for Barred Galaxies with Inner Disks}
    \label{tab:galaxies-id}
    \begin{tabular}{llrrrrrrlr}
\hline
Name & Type (RC3) & $B_{t,c}$ &  $R_{25}$ & $i$     & Disk PA & Distance
     & \sigzero{} & Rot & Notes \\
     &            &           & \arcsec{} & \degr{} & \degr{} &   Mpc
     &  \kms{}    &     &       \\
(1)  & (2)        & (3)       & (4)       & (5)     & (6)     & (7)
     & (8)        & (9) & (10) \\
\hline
 NGC 151 &                SB(r)bc & 11.65 &  111 &   65 &   75 &  48.9 & \ldots  & + & \\
 NGC 470 &                SA(rs)b & 11.99 &   84 &   56 &  155 &  31.3 &  143 & + & \\
NGC 1398 &     (\ronetwo)SB(rs)ab & 10.29 &  212 &   45 &   96 &  16.1 &  205 & + & \\
NGC 2787 &             SB(r)$0^+$ & 11.27 &   95 &   55 &  109 &   7.5 &  195 & + & \\
NGC 2880 &                SB$0^-$ & 12.32 &   62 &   52 &  144 &  21.9 &  142 & $-$? & \\
NGC 3266 &              SAB$0^0$? & 13.58 &   46 &   35 &   85 &  25.1 &  116 & \ldots  & 1\\
NGC 3384 &            SB(s)$0^0$: & 10.68 &  165 &   61 &   50 &  11.6 &  140 & \ldots  & \\
NGC 3412 &             SB(s)$0^0$ & 11.29 &  109 &   52 &  153 &  11.3 &  108 & \ldots  & \\
NGC 4143 &            SAB(s)$0^0$ & 11.70 &   69 &   59 &  144 &  15.9 &  270 & $-$ & \\
NGC 4262 &            SB(s)$0^-$? & 12.26 &   56 &   29 &  153 &  15.3 &  188 & \ldots  & \\
NGC 4386 &              SAB$0^0$: & 12.47 &   74 &   48 &  140 &  27.0 &  197 & \ldots  & 1\\
NGC 4612 &            (R)SAB$0^0$ & 12.08 &   74 &   44 &  143 &  15.3 &   61 & \ldots  & \\
NGC 4754 &            SB(r)$0^-$: & 11.40 &  137 &   63 &   23 &  16.8 &  200 & \ldots  & 1\\
NGC 7007 &               SA$0^-$: & 12.90 &   57 &   55 &    2 &  37.6 &  145 & \ldots  & \\
UGC 6062 &            SAB(r)$0^0$ & 13.29 &   36 &   55 &   25 &  35.7 &  178 & \ldots  & 1\\
ESO 378-G020 &           SB(rs)$0^0$? & 13.17 &   39 &   57 &   34 &  38.3 &  132 & \ldots  & 1\\
ESO 443-G017 &           (R)SB(rl)0/a & 13.53 &   42 &   50 &   23 &  39.3 & \ldots  & $-$ & \\
\hline
    \end{tabular}

\medskip

Columns are the same as for Table~\ref{tab:galaxies-db}.  As noted in
Table~\ref{tab:galaxies-db}, the double-barred galaxies NGC~3368,
3945, 4785, and 7187 also have inner disks; to save space, they are 
not repeated here.  Notes for individual galaxies: 1 = unsharp masking
suggests inner disk is probably an inner bar.  Sources of data for
individual galaxies are discussed in Sect.~\ref{sec:notes}.
\end{minipage}
\end{table*}

\section{Notes on Individual Galaxies}
\label{sec:notes}

For each galaxy, I first list the ``source'' -- i.e., the citation(s)
where the double-bar identification was first made, along with
significant early studies of the double-bar/inner-disk system as such
-- and then briefly discuss where the bar, disk, and other measurements
come from.  Unless otherwise noted, outer-disk position angle and
inclination are taken from the 25th-magnitude values in
\citet[][hereafter RC3]{rc3}\footnote{Inclination is derived from the
axis ratio $b/a$, assuming an intrinsic disk thickness $c/a = 0.2$.},
and distances are based on the redshift listed in LEDA, corrected for
Virgocentric motion \citep{paturel97}, and assuming a Hubble constant
$H_{0} = 75$ \kms{} Mpc$^{-1}$.  Similarly, central velocity dispersion
\sigzero{} is from the compilation of \citet{mcelroy95}, unless
otherwise noted.

\textbf{NGC 151 (inner disk)}: \citet{marquez99}, from which
the bar measurements were taken.  Inner-disk measurements are from a
NICMOS3 F160W image, and agree well with the M\'arquez et al.\ values.

\textbf{NGC 357}: \citet{mrk97}.  Outer-bar measurements are
from the publically available $K$-band image of Mulchaey et al.;
inner-bar measurements are from a NICMOS2 F160W image.  Outer disk
inclination and PA are from ellipse fits to an unpublished, deep $I$-band
image by Alfonso L\'opez Aguerri and Enrico Maria Corsini.

\textbf{NGC 470 (inner disk)}: \citet{w95} and \citet{friedli96a}. 
Bar and inner-disk measurements are based on the ellipse fits of
\citet{friedli96a}; \sigzero{} is from \citet{prugniel01}.

\textbf{NGC 718}: \citet{erwin00}; see \citet{erwin-sparke03}. 
Adopted inner-bar \lbar{} is that of the blue nuclear ring in
\citet{erwin-sparke03}.

\textbf{NGC 1068 (M77)}: \citet{eva00}.  Depending on how one counts,
as many as \textit{five} separate bars have been suggested for this
galaxy.  The bar universally agreed upon is that first found in the
near-IR by \citet{scoville88}, with $a \sim 17\arcsec$.  \citet{eva00}
presented both imaging and kinematic evidence for a much larger,
rounder bar outside; this is the same as the ``oval disk'' discussed by
\citet{kormendy-norman79}.  This galaxy is thus similar to NGC~4736
(below): a fairly strong and large inner bar resides inside a very
large, weak outer bar/oval disk.  The measurements presented here are
based on the images and ellipse fits of \citet{alonso-herrero98} for
the inner bar (with \lbar{} based on the size of the nuclear
pseudo-ring surrounding it) and on the 2MASS $J$-band image for the
outer bar.  Distance, inclination, and outer-disk PA are from
Schinnerer et al.\ and references therein.

Other studies have proposed additional, smaller bars for this galaxy,
but the evidence for these is ambiguous or dubious.  \citet{rouan98},
using adaptive-optics, near-IR images and archival F547M WFPC2 images,
suggested no fewer than \textit{three} nested bars, all smaller than
the well-known 17\arcsec{} bar (i.e., the inner bar in this catalog). 
At least some of these features are probably due to dust and scattered
nuclear radiation, and possibly the radio jet as well
\citep[][]{weinberger99,bock00}.  Finally, \citet{laine02} reported
two bars: the well-known IR bar, and a smaller bar with deprojected $a
= 1.7\arcsec$, possibly matching the middle of Rouan et al.'s three
bars.  But Schinnerer et al.\ found no evidence for a nuclear bar on
those size scales, and argued instead for a warped molecular disk,
which might explain some of the features seen by Rouan et al.\ and
Laine et al.

\textbf{NGC 1097}: \citet{shaw93} and \citet{bc93}.  Outer-bar
measurements are from the 2MASS $H$-band image, except that $\emax$ is
from the $I$-band ellipse fits of \citet{w95}, which are higher
resolution, and \lbar{} is from spiral arms trailing off the ends of
the bar, as seen in the 7$\mu$m ISO image \citep[][available via
NED]{roussel01}.  Inner-bar measurements are based on those of
\citet{quillen95} and the ellipse fits in \citet{shaw93} and
\citet{friedli96a}.  Outer-disk PA and inclination are from \hi{} and
emission-line kinematics \citep{ondrechen89,storchi-bergman96}.

\textbf{NGC 1241}: \citet{laine02}.  Outer-bar measurements
are from a NICMOS3 F160W image, with inner-bar measurements from a
NICMOS2 F160W image; the inner-bar \amax{} is based on an extremum in
the ellipse-fit PA, and \lbar{} is from the size of the nuclear ring.

\textbf{NGC 1291}: \citet{dev74,dev75}.  Outer-bar
measurements are from the DSS2 red image, except for bar PA from
\citet{csrg}.  Inner-bar measurements are from an $I$-band image
(courtesy Oak-Kyung Park). Disk inclination is from \hi{} kinematics
\citep{vandriel88}; the disk PA is undefined.

\textbf{NGC 1317}: \citet{schweizer80}.  Outer-bar measurements are
from the publically available $R$-band image of \citet{marcum01};
inner-bar measurements are from the publically available $K$-band image
of \citet{mrk97}, except that \lbar{} is from size of the inner \ha{}
nuclear ring.  The outer-disk inclination is from the axis ratio quoted
in \citet{bc93}; the distance is a default value for the Fornax
Cluster, from the \textit{HST} Key Project study of \citet{freedman01}.

\textbf{NGC 1398 (inner disk)}: \citet{erwin04}, based partly
on \citet{jungwiert97}.  The inner ``triaxial bulge'' identified by
Jungwiert et al.\ \citep[see also][]{w95}, with $\amax \approx
13\arcsec$, appears to be a large stellar nuclear ring; the inner disk
listed here shows up in \textit{HST} images.  Bar measurements are from the
ellipse fits of \citet{w95} and \citet{jungwiert97}; inner-disk
measurements are from WFPC2 F606W and F814W images.  Outer-disk
inclination and PA are from \hi{} kinematics \citep{moore95}.

\textbf{NGC 1433}: \citet{sandage79} and \citet{buta86}. 
Outer-bar measurements are from the ellipse fits of \citet{w95} and
\citet{buta01}, with \lbar{} from the size of the inner ring in Buta
et al.  Inner-bar measurements are from the ellipse-fits of
\citet{jungwiert97}, with \lbar{} from size of nuclear ring (Buta et
al.).  Outer-disk inclination and PA are from \hi{} kinematics
\citep{ryder96}.

\textbf{NGC 1543}: \citet{dev75} and \citet{sandage79}. 
Outer-bar measurements are from a DSS image; inner-bar measurements
are from WFPC2 F814W images, with \lbar{} from the size of the stellar
nuclear ring.  Outer disk inclination is from the axis ratio quoted in
\citet{bc93}; distance is from \citet[][surface-brightness
fluctuation]{tonry01}.  See Fig.~\ref{fig:umask-innerbars}.

\textbf{NGC 1808}: \citet{kotilainen96}, \citet{tacconi-garman96}, and
\citet{jungwiert97}.  Outer-bar measurements are from publically
available $H$-band image of the OSU Bright Spiral Galaxy Survey
\citep{eskridge02}; inner-bar measurements are from the ellipse fits
of \citet{jungwiert97}.  Outer-disk inclination is based on the axis
ratio of the outer spiral arms listed in \citet{koribalski93}.

The outer bar in this galaxy is quite peculiar (similar to those of
NGC~4725 and ESO 443-39), and its position angle is somewhat
ill-defined.

\textbf{NGC 2217}: \citet{jungwiert97}.  Measurements for both
bars are from the ellipse fits of Jungwiert et al., except that
\lbar{} for outer bar is from \citet{ohta90}, based on the decline of
the bar-interbar intensity ratio\footnote{\citet{erwin-sparke03} found
that the Ohta et al.\ measurements were a good match to their
definition of \lbar, and the ellipse fits for this galaxy do not
provide useful constraints.}.  Outer-disk inclination and PA are from
stellar kinematics \citep{bettoni90}.

\textbf{NGC 2642}: \citet{erwin04}.  Bar measurements are from
a NICMOS3 F160W image, with \lbar{} for inner bar from size of the
nuclear ring and \lbar{} for outer bar from size of the inner
ring/spiral arms encircling the bar.

\textbf{NGC 2646}: \citet{erwin04}.  Bar measurements are from
ellipse fits to an $R$-band image.  Disk PA is unknown; RC3 gives
nearly circular outer isophotes.

\textbf{NGC 2681}: \citet{w95} and \citet{erwin99}.  See
\citet{erwin-sparke03}.

\textbf{NGC 2787 (inner disk)}: \citet{erwin00}; see
\citet{erwin-sparke03}.  The inner disk is discussed in greater detail
in \citet{erwin03-id}.

\textbf{NGC 2859}: \citet{kormendy79,kormendy82a} and
\citet{w95}.  See \citet{erwin-sparke03}.

\textbf{NGC 2880 (inner disk)}: \citet{erwin00}; see
\citet{erwin-sparke03}.

\textbf{NGC 2950}: \citet{kormendy79,kormendy82a} and
\citet{w95}.  See \citet{erwin-sparke03}.

\textbf{NGC 2962}: \citet{erwin00}; see \citet{erwin-sparke03}.

\textbf{NGC 3081}: \citet{buta90a}.  Outer-bar measurements
are based on those of \citet{friedli96a}, \citet{mrk97}, and
\citet{buta-purcell98}; \lbar{} is from the size of the inner ring,
measured in the publically available $K$-band image of Mulchaey et al. 
Measurements for inner bar are from a NICMOS2 F160W image; \lbar{} is
from the size of the nuclear ring.  Disk inclination and PA are from
kinematic and photometric arguments in Buta \& Purcell.

\textbf{NGC 3266 (inner disk)}: \citet{erwin04}.  Bar
measurements are from WFPC2 F702W images \citep{rest01}; the outer
disk orientation is from ellipse fits to the WFPC2 mosaic image, and
\sigzero{} is from \citet{wegner03}.  Unsharp masking suggests that
the inner disk is probably an inner bar.

\textbf{NGC 3275}: \citet{mrk97}.  Outer-bar measurements are from the
publically available images of Mulchaey et al.; inner-bar measurements
are from a NICMOS3 F160W image, with \lbar{} based on the nuclear ring
size.

\textbf{NGC 3358}: \citet{bc91,bc93}.  Bar measurements are
from ellipse fits to $V$- and $I$-band images (outer bar \amax{} is
from PA minimum), with outer disk position angle from ellipse fits to
a $B$-band image, all kindly provided by Ron Buta.  Outer disk 
inclination is from the axis ratio quoted in \citet{bc93}.

\textbf{NGC 3368 (M96; double bar + inner disk)}: Source:
\citet{jungwiert97}.  The $H$-band image of Jungwiert et al.\ is not
large enough to fully cover the outer bar; thus, the ``double bar''
they reported is actually the inner bar + an intermediate elliptical
structure, identified here as an inner disk due to its low ellipticity
and alignment with the outer disk.  Outer-bar, inner-disk, and
inner-bar measurements are from the publically available $K$-band image
of \citet{mh01}, except that inner-bar \amax{} is based on ellipse fits
to a NICMOS2 F160W image and \lbar{} on the size of the nuclear ring.
See \citet{erwin03-bigbars} for a discussion of the outer-disk
orientation; distance is from Cepheid measurements \citep{freedman01}.

\textbf{NGC 3384 (inner disk)}: \citet{busarello96}.  Outer-bar
measurements are from the ellipse fits of Busarello et al.; $\amax$ in
this case is actually the ellipticity \textit{minimum}, which is also
the extremum in the PA twist.  Bar PA is from analysis of Busarello et
al., which showed that the best ellipse-fit PA for the bar is $\sim
60\degr$ away from the bar's true position angle.  (The problems with
the ellipse fits comes from the fact that this galaxy is highly
inclined, and the bar is very close to the projected minor axis.)
Inner disk measurements are from WFPC2 F814W and NICMOS2 F160W images.
Distance is from \citet[][surface-brightness fluctuation]{tonry01}, and
\sigzero{} is the average of several recent measurements
\citep{fisher97,neistein99,gebhardt03}.

\textbf{NGC 3393}: \citet{jungwiert97}, \citet{alonso-herrero98}, and
\citet{greusard00}.  Bar measurements are based the publically
available $K$-band image of \citet{mrk97}, plus the near-IR ellipse
fits of Jungwiert et al.\ and Greusard et al.  Outer-disk inclination
is from \citet{kornreich98}; outer-disk PA is undefined due to the low
inclination.

\textbf{NGC 3412 (inner disk)}: \citet{erwin00}; see \citet{erwin-sparke03}.

\textbf{NGC 3941}: \citet{erwin00}; see \citet{erwin-sparke03}.

\textbf{NGC 3945 (double bar + inner disk)}:
\citet{kormendy79,kormendy82a} and \citet{w95}.  See \citet{erwin99}
and \citet{erwin-sparke03}; for a detailed study of the inner disk, 
see \citet{erwin03-id}.

\citet{w95} suggested this galaxy might be \textit{triply} barred
\citep[it is listed as such by][]{moiseev01}; however, \citet{erwin99}
used \textit{HST} images to show that it was double-barred, with a
large inner disk dominating the region between the two bars.  Similar
bar-disk-bar systems are NGC~3368, NGC~4785, and NGC~7187.

\textbf{NGC 4143 (inner disk)}: \citet{erwin00}; see
\citet{erwin-sparke03}.  Central velocity dispersion is from
\citet{dinella95}.

\textbf{NGC 4262 (inner disk)}: \citet{shaw95}.  See \citet{erwin04}
for source of bar and disk measurements.  The distance is a default
value for the Virgo Cluster, from the \textit{HST} Key Project study
of \citet{freedman01}.

\textbf{NGC 4303 (M61)}: \citet{colina-wada00}.  Outer-bar measurements
are from the publically available $K$-band image of \citet{mh01}, with
\lbar{} from a spiral arm crossing over the southern end of the bar;
inner-bar measurements are from a NICMOS2 F160W image.  Outer-disk
inclination and PA are from the kinematic arguments of \citet{eva01};
the distance is the default Virgo Cluster distance (see note for
NGC~4262).

The outer-bar \amax{} is based on a minimum in the PA, since there is
no clear ellipticity maximum within the bar; the ellipticity maximum
at $a \approx 45\arcsec$ \citep[e.g.,][]{laine02} is due to spiral
arms outside the bar.

\textbf{NGC 4314}: \citet{benedict93}.  See \citet{erwin-sparke03},
and \citet{erwin03-bigbars} for measurements of the outer bar;
\sigzero{} is from \citet{barth02}.

\textbf{NGC 4321 (M100)}: \citet{pierce86}, \citet{shaw95},
\citet{knapen95a}.  Outer-bar measurements are from ellipse fits to
the 2MASS $K$-band image, with \lbar{} from spiral arms crossing the
ends of the bar; inner-bar measurements are from the near-IR ellipse
fits in \citet{perez-r00}.  Outer-disk inclination and PA are from the
\hi{} kinematics of \citet{knapen93}; distance is from Cepheids
\citep{freedman01}.

\textbf{NGC 4340}: \citet{kormendy79} and \citet{w95}.  Bar and
outer-disk measurements are from an MDM $R$-band image obtained by Paul
Schechter; inner-bar \lbar{} is from size of stellar nuclear ring; see
Fig.~\ref{fig:umask-innerbars} and \citet{erwin01-lapalma}.  The
distance is the default Virgo Cluster distance (see note for NGC~4262).

\textbf{NGC 4386 (inner disk)}: \citet{erwin00}; see
\citet{erwin-sparke03}. Unsharp masking suggests that
the inner disk is probably an inner bar.

\textbf{NGC 4503}: \citet{erwin04}.  Bar measurements are from
the $H$-band image of \citet{jungwiert97}, courtesy Fran\c{c}oise
Combes, for the outer bar and from a WFPC2 F702W image for the inner
bar.  The distance is the default for the Virgo Cluster (see note for
NGC~4262).

\textbf{NGC 4612 (inner disk)}: \citet{erwin04}, based on
\citet{jungwiert97}.  All measurements, including outer-disk
orientation, are from an MDM $R$-band image, courtesy Paul Schechter;
the outer-bar and outer-disk measurements agree well with measurements
made using the $R$-band image of \citet{frei96}, available via NED.  
Central velocity dispersion is from \citet{wegner03}.

\textbf{NGC 4725}: \citet{laine02}.  Outer-bar measurements are from
the $r$-band image of \citet{frei96} and a $K$-band image kindly
provided by Johan Knapen, with \lbar{} from the inner ring; the outer
bar is rather peculiar (similar to those of NGC~1808 and ESO 443-39),
and its position angle is not well defined.  Inner-bar measurements are
from a NICMOS2 F160W image.  See \citet{erwin03-bigbars} for a
discussion of the outer-disk orientation; the distance is from Cepheid
measurements \citep{freedman01}.

\textbf{NGC 4736 (M94)}: \citet{shaw93} and \citet{moellenhoff95}.  As
with NGC~1068, the outer bar is rather weak, and has been termed an
``oval disk'' rather than a bar.  However, there is ample kinematic
evidence that it rotates and behaves dynamically like a bar
\citep[e.g.,][and references therein]{moellenhoff95,vandriel96}.  Outer
disk PA is from the kinematics and modeling of M\"ollenhoff et al.\ and
van Driel et al., with inclination being the average of photometric and
kinematic inclinations from those studies and the \hi{} observations of
\citet{mulder93}.  Distance is from \citet[][surface-brightness
fluctuation]{tonry01}.

\textbf{NGC 4785 (double bar + inner disk)}: \citet{marquez99},
\citet{erwin04}.  Bar and inner disk measurements are from a NICMOS2
F160W image and the ellipse fits of \citet{marquez99}; outer-bar
\lbar{} is from spiral arms/inner ring.  I classify the ``inner bar''
of M\'arquez et al.\ as an inner disk; the inner bar listed here was
noted by them as an ``inner elongation'' in their unsharp mask of the
NICMOS image.  The outer bar is somewhat weak and difficult to
distinguish from the spiral arms outside it.  The central velocity
dispersion is from \citet{oliva99}.

\textbf{NGC 4754 (inner disk)}: \citet{kormendy79, shaw95}.  All
measurements, including outer-disk orientation, are from WIYN $R$-band
images \citep{erwin04}; distance is from \citet[][surface-brightness
fluctuation]{tonry01}.  Unsharp masking suggests that the inner disk
is probably an inner bar.

\textbf{NGC 4984}: \citet{jungwiert97}.  Bar measurements are from the
ellipse fits and isophotes in Jungwiert et al., with \lbar{} for the
outer bar from the inner-ring size listed in \citet{bc93}.  Because
the galaxy has prominent inner and outer rings (the RC3 position angle
is basically that of the inner ring/lens), the outer disk orientation
is based on considerations of typical inner- and outer-ring
orientations and axis ratios compared with those of the galaxy's
rings, similar to the approach used in \citet{erwin99,erwin-sparke03}
for other galaxies with strong outer rings.

\textbf{NGC 5365}: \citet{mrk97}.  Bar measurements are from
ellipse fits to the publically available $K$-band image of Mulchaey et
al, with \amax{} for the inner bar from an extremum in the PA.  Outer 
disk orientation is from LEDA.

\textbf{NGC 5728}: \citet{shaw93} and \citet{bc93}.  Outer-bar
measurements are from the ellipse fits of \citet{w95} and
\citet{prada99}; inner-bar measurements are from the near-IR ellipse
fits and isophotes of \citet{shaw93}, except that \amax{} and \emax{}
are from the higher-resolution $I$-band ellipse fits of Prada \&
Guti\'{e}rrez.  The outer disk orientation is based on discussion in
\citet{nagar99}.

\textbf{NGC 5850}: \citet{bc93}.  Bar measurements are from a Nordic
Optical Telescope $I$-band image \citep[from the BARS
Project;][]{lourenso01} and the near-IR ellipse fits of
\citet{friedli96a}.  \lbar{} is from the inner ring \citep{prieto97}
for the outer bar and from the nuclear ring for the inner bar.  Outer
disk orientation is from the \hi{} kinematic study of
\citet{higdon98}.

\textbf{NGC 6654}: \citet{erwin00}; see \citet{erwin-sparke03}.

\textbf{NGC 6684}: \citet{phillips96}.  Outer-bar measurements are from
a WFPC2 F814W mosaic image, with \lbar{} from the size of the inner
pseudo-ring; inner-bar measurements are from a NICMOS3 F160W image.
Outer disk inclination and PA are from ellipse fits to an unpublished,
deep $I$-band image by Alfonso L\'{o}pez Aguerri and Enrico Maria Corsini;
distance is from \citet[][surface-brightness fluctuation]{tonry01}.
See Fig.~\ref{fig:umask-innerbars}.

\textbf{NGC 6782}: \citet{bc93} and \citet{w95}.  Outer-bar
measurements are based on the ellipse fits of \citet{friedli96a},
\citet{jungwiert97}, and a WFPC2 F814W image; inner-bar measurements
are from the F814W image (good agreement with ellipse-fit values from
Friedli et al.\ and Jungwiert et al.).  Outer-disk inclination and PA
are from the isophotal analysis in \citet{quillen97}; \sigzero{} is
from \citet{idiart96}.

\textbf{NGC 7007 (inner disk)}: \citet{mrk97}.  Bar and
inner-disk measurements are from the publically available $K$-band
image of Mulchaey et al; \sigzero{} is from \citet{wegner03}.

\textbf{NGC 7098}: Source \citet{bc93} and \citet{w95}.  Outer-bar
measurements are from the ellipse fits and isophotes of
\citet{friedli96a}; inner-bar measurements are from a WFPC2 F814W image
(good agreement with ellipse-fit values from Friedli et al.). 
Outer-disk inclination is from the axis ratio quoted in \citet{bc93}.

\textbf{NGC 7187 (double bar + inner disk)}: \citet{buta90b} and
\citet{w95}.  Outer-disk inclination and PA are from Buta; bar
measurements are based on those of Buta and \citet{w95}.  Adopted
outer-bar length is that of the inner ring surrounding it.

\citet{w95} identified this a \textit{triple}-barred galaxy; however,
the middle ``bar'' is low enough in ellipticity and close enough in
alignment to the outer disk to qualify as an inner disk, making this
galaxy similar to NGC~3368, NGC~3945, and NGC~4785.

\textbf{NGC 7280}: \citet{erwin00}; see \citet{erwin-sparke03}.

\textbf{NGC 7716}: \citet{laine02}.  Outer-bar measurements are from
the publically available $K$-band image of \citet{mrk97}; inner-bar
measurements are from a NICMOS2 F160W image.

\textbf{Mrk 573}: \citet{capetti96}, \citet{alonso-herrero98}. 
\citet{laine02} suggested that this galaxy was \textit{triple}-barred;
however, the largest of their three bars, identified via ellipse fits,
is actually an outer ring \citep[e.g.,][]{afanasiev96}.  Outer-bar
measurements are from the $J$-band ellipse fits of
\citet{alonso-herrero98}, with inner-bar measurements from a NICMOS1
F160W image; \lbar{} for the inner bar is from the nuclear ring.

\textbf{UGC 524}: \citet{pogge02}.  Bar measurements are from a WFPC2
F814W image.  Outer-disk inclination and PA are from LEDA.

\textbf{UGC 6062 (inner disk)}: \citet{rest01}.  Based on ellipse fits
to the WFPC2 F702W images; bar and inner-disk position angles are from
inspection of the isophotes and unsharp masks.  The outer disk
inclination and PA are from LEDA. Unsharp masking suggests that the
inner disk is probably an inner bar.

\textbf{ESO 215-31}: \citet{greusard00}.  Bar measurements are
from the near-IR ellipse fits of Greusard et al.

\textbf{ESO 320-30}: \citet{greusard00}.  Bar measurements are
from the near-IR ellipse fits of Greusard et al.

\textbf{ESO 378-20 (inner disk)}: \citet{rest01}.  Bar and
inner disk measurements are from WFPC2 F702W images, with outer disk
orientation from ellipse fits to the WFPC2 mosaic image; \sigzero{} is
from \citet{wegner03}.  Unsharp masking suggests that the inner disk
is probably an inner bar.

\textbf{ESO 443-17 (inner disk)}: \citet{greusard00}.  Bar
measurements are from the near-IR ellipse and isophotes of Greusard et
al.

\textbf{ESO 443-39}: \citet{rest01}.  Bar measurements are from WFPC2
F702W images; the outer bar is peculiar and difficult to measure
accurately (similar to NGC~1808 and NGC~4725).  Central velocity
dispersion is from \citet{wegner03}.

\textbf{ESO 447-30}: \citet{rest01}.  Bar measurements are 
from WFPC2 F702W images; \sigzero{} is from \citet{wegner03}.  The 
outer disk orientation is somewhat uncertain, since $D_{25}$ coincides 
with an outer ring; however, all three rings (nuclear, inner, and 
outer) share a common PA of 34--36\degr{} and ellipticities of 
0.32--0.37.

\textbf{IRAS 03565+2139}: \citet{erwin04}.  Based on $R$-band
observations by Chris Conselice and Jay Gallagher with the 3.5 m WIYN
Telescope.  \lbar{} values for the inner and outer bars are from the
sizes of the nuclear and inner rings, respectively.  The inclination
is based on inverting the $H$-band Tully-Fisher relation \citep[as
given in ][p.~425]{bm98}, using the $H$-band magnitude and the \hi{}
width $W_{20}$ from \citet{vandriel95}.  Note that van Driel et al.\
considered this to be a deviant from their T-F relation, but that
conclusion was based on the assumption that the RC3 axis ratio was
that of the outer disk; the $R$-band image shows that the RC3 axis
ratio is actually that of the inner ring, which is most likely not 
intrinsically circular \citep[e.g.,][]{buta86,buta95}.

\section{Unconfirmed or Ambiguous Inner Bars and Disks}
\label{sec:ambiguous}

Here, I list galaxies for which the evidence for a double-bar or
inner-disk system is ambiguous or uncertain.  Better observations
(typically, near-IR and/or higher resolution) are needed in order to
make definitive judgments one way or the other.  This also includes
some galaxies which are probably double-barred, but for which useful
measurements are lacking.

\textbf{NGC 613}: Suggested as double-barred by \citet{jungwiert97},
using their $H$-band image.  Inspection of WFPC2 images reveals a
pronounced and very elliptical nuclear ring with strong star formation,
having the same size, ellipticity, and orientation as the suggested
inner bar.  While unsharp masking of the $H$-band image (courtesy
Fran\c{c}oise Combes) displays something like an inner bar feature, the
bright ends of the ``bar'' are resolved into multiple star clusters in
the \textit{HST} images.  This indicates that the feature is probably
\textit{not} an inner bar, but higher-resolution near-IR (preferably
$K$-band) images are needed to decide one way or the other.

\textbf{NGC 1079}: Listed by \citet{moiseev01}, on the basis of an
ellipse-fit feature in \citet{jungwiert97}.  However, the ellipticity
peak is very weak, and Jungwiert et al.\ preferred to attribute it to
``bulge triaxiality.''

\textbf{NGC 1326}: This is one of two double-barred galaxies
originally discussed by \citet{dev74}; it has also been reported by
\citet{bc93} and \citet{w95}.  Inspection of WFPC2 F814W images shows
very strong dust lanes inside a dusty, star-forming nuclear ring,
which combine to produce the illusion of an inner bar.  Near-IR images
are needed to determine if there might still be an inner bar inside
the nuclear ring.

\textbf{NGC 1353}: \citet{jungwiert97} noted a possible inner bar,
though they were doubtful given the galaxy's high inclination($i \ap
70\degr$).  Based on its orientation, the ellipse-fit peak noted by
Jungwiert et al.\ is a candidate inner disk.  Inspection of a WFPC2
F606W image shows a very dusty nuclear ring with $r \ap 5\arcsec$ and
complex dust lanes inside; this raises the possibility that residual
dust extinction could cause the ellipse peak seen in $H$.  Higher
resolution near-IR imaging is needed to determine what is responsible
for the ellipticity peak.
  
\textbf{NGC 1371}: Listed by \citet{moiseev01}, with a reference to
\citet{w95}; however, this galaxy is not in the latter study.  Since
the bar sizes listed by Moiseev match those reported by Wozniak et
al.\ for NGC 1317, this is probably an accidental
transposition/duplication.

\textbf{NGC 1415}: \citet{garcia-b00} reported an inner ``stellar
bar'' in red continuum images of this barred Sa.  Since the position
angle they report ($\sim 150\degr$) is very close to the RC3
position angle (148\degr), this is a candidate inner disk.  No
archival images are available; 2MASS images also suggest an inner
elliptical feature, but there are indications of isophote twists, so
this might be a nuclear spiral instead.  Higher-resolution near-IR
images are needed.

\textbf{NGC 2339}: Suggested as double-barred by \citet{laine02}. 
Inspection of WFPC2 F606W and NICMOS2 F110W and F160W images
indicates that the nuclear region has both strong dust lanes and
star formation; the ``bar'' detected by Laine et al.\ appears to be
an IR-bright (star-forming?)  ring, with at least two bright nuclei
inside.  Higher-resolution $K$-band images are probably needed to be
certain there is no inner bar here.

\textbf{NGC 2639}: Suggested as double-barred by \citet{marquez99},
though \citet{laine02} found \textit{no} bars in their analysis of
this galaxy.  Inspection, ellipse fitting, and unsharp masking of the
NICMOS2 and NICMOS3 F160W and WFPC2 F547M images shows some evidence
for a weak bar with $a \sim 4$--5\arcsec{} (note that this does not
match either of M\'arquez et al.'s proposed bars), oriented almost
parallel with the projected outer disk; its presence is primarily
signaled by the appearance of two classic, curved leading-edge dust
lanes and a weak feature in the $H$-band ellipse fits.  These dust
lanes curve together at $r \la 1.5\arcsec$ and meet in a chaotic,
dusty region.  There is tenuous evidence for a very weak elliptical
feature with $a \sim 1.0\arcsec$ in the $H$-band ellipse fits, but
there is also some indication that the dust seen in the WFPC2 image is
distorting the $H$-band isophotes.

\textbf{NGC 2442}: Suggested by \citet{moiseev01}, based on the
suggestion of a triaxial bulge by \citet{baumgart86};
\citet{jungwiert97} list this in their ``twisted isophotes'' category. 
Inspection of their $H$-band isophotes suggests there \textit{may} be
a nuclear bar with $a \sim 2.5$--3\arcsec; higher-resolution
observations are needed for confirmation and accurate measurements. 
The outer bar is peculiar and difficult to define; it is somewhat
similar to those of NGC~1808 and NGC~4725.

\textbf{NGC 2685}: \citet{erwin02,erwin-sparke03} listed this among
their inner-disk galaxies.  While the inner disk identification is
strong, identification of the ``spindle'' as a bar (rather than a
highly inclined large-scale disk) is sufficiently uncertain that
this should probably not be considered a barred galaxy \citep[see
the discussion in ][]{erwin-sparke03}.

\textbf{NGC 2782}: Suggested by \citet{jogee99}.  As they point out, 
the evidence for a nuclear bar is rather good, but the case for an 
outer bar (or outer oval, in this case) is uncertain.

\textbf{NGC 2811}: Suggested by \citet{marquez99}.  The near-IR 
ellipse fits of \citet{jungwiert97} indicate a possible inner 
ellipticity peak, though it is at the limits of their resolution.  
Given the alignment, it is a candidate inner disk, but without higher 
resolution images the detection remains dubious.

\textbf{NGC 2935}: Suggested by \citet{jungwiert97}.  Because
\citet{bc93} reported a nuclear ring in this galaxy, there is the
possibility that the ``inner bar'' detection is due to, or affected
by, the nuclear ring.  This is a promising candidate, but no higher
resolution images are available.

\textbf{NGC 3504}: \citet{perez-r00} noted a possible inner bar in 
this SBb galaxy, but also an apparent double nucleus.  This latter 
feature makes it difficult to be certain about the inner bar.

\textbf{NGC 4253}: Suggested as double-barred by \citet{marquez99},
based on ellipse fits.  Though these are at the probable limit of
their resolution, they are plausible; however, the ellipse fits of
\citet{peletier99} do not agree with this.  Unfortunately, the NICMOS2
F160W image is marred by saturation and a very strong diffraction
pattern from the Seyfert nucleus.  The inner isophotes appear fairly
round and aligned with the (outer?)  bar.  The diffraction spikes are
less severe in the WFPC2 F606W image, where there is no obvious sign
of an inner bar.  Strong dust lanes visible in the optical image could
be the cause of the ellipse fit variations seen by \citet{marquez99};
they also prevent a clear determination using the optical image.

\textbf{NGC 4274}: Suggested as double-barred by \citet{shaw95}. 
Their near-IR image strongly suggests an inner bar, almost
perpendicular to the outer bar, but no measurements are available. 
Unfortunately, the available WFPC2 images (F555W) show that the
central region is extremely dusty, so near-IR images are required for
measurement of the inner bar.

\textbf{NGC 4290}: Suggested by \citet{marquez99}, though they 
labeled it as ``uncertain.''  Their near-IR ellipse fits show an 
ellipticity peak, but without a stationary position angle, so a 
nuclear spiral is a possiblity.  Higher-resolution images are needed.

\textbf{NGC 4594 (M014)}: \citet{emsellem00} suggested that the
Sombrero Galaxy might be a double-bar system, based on color maps and
2D kinematics.  The extremely high inclination of this galaxy makes
finding (and measuring) any bar difficult, and Emsellem \& Ferruit
point to alternate explanations and emphasize the tentative nature of
this identification.  (Part of the evidence for the inner bar is an
apparent straight dust lane, evoking the classic leading-edge dust
lanes of large-scale bars; but \citet{witold02} and \citet{shlosman02}
argue, on the basis of hydrodynamical simulations of double bars, that
straight dust lanes are probably \textit{not} characteristic of inner
bars.)

\textbf{NGC 4750}: Suggested as double-barred by \citet{laine02}. 
Inspection of NICMOS2 images show a clear, strong nuclear bar;
however, the presence of the \textit{outer} bar is much less certain. 
Optical images (including WFPC2 F606W) are too dusty to be useful. 
2MASS near-IR images show an elliptical region corresponding to the $a
= 14\arcsec$ outer bar of Laine et al; however, unsharp masking shows
that at least part of this structure is a tightly wrapped
spiral/pseudo-ring.  Higher-resolution and higher S/N near-IR images
are needed to determine if a bar is present in addition to the
spirals, and to properly measure its size and orientation.

\textbf{NGC 4941}: \citet{greusard00} found a nuclear bar in their
near-IR images of this SABab galaxy, but argued that there was no
outer bar.  On the other hand, \citet{eskridge02} classify the galaxy
as SAB based on their low-resolution near-IR images, and
\citet{kormendy82b} includes it in his list of galaxies with ``oval
disks,'' so it may be similar to NGC~1068 and NGC~4736: a galaxy with
a strong inner bar and a weak outer bar.  For the time being, the
evidence for an outer bar remains too tenuous to classify this as a
double-barred galaxy.

\textbf{NGC 5101}: Suggested as a possible double bar by
\citet{jungwiert97}.  However, as they pointed out, the ellipticity
peak is very weak and close to their seeing limit.  Unsharp masking of
their $H$-band image (courtesy Fran\c{c}oise Combes) suggests a
possible nuclear ring or spiral.  No higher-resolution images are
available.

\textbf{NGC 5566}: Suggested as double-barred by \citet{jungwiert97};
the orientation of inner structure would make it an inner disk. 
However, unsharp masking of WFPC2 F606W images suggests that the inner
elliptical feature is a pair of spiral arms.  Due to the presence of
numerous dust lanes, this is an uncertain classification;
high-resolution near-IR images are needed to be certain one way or the
other.

\textbf{NGC 5905}: Suggested as double-barred by \citet{w95}; the
proposed inner bar is near the limits of their resolution, and could
be due to the nuclear ring they also report.  Higher-resolution
images, preferably near-IR, are needed.

\textbf{NGC 6155}: Suggested as a possble double bar by
\citet{marquez99}.  In this case, it is the outer bar whose existence
is in doubt; M\'arquez et al.\ note that it could be spirals arms
instead.

\textbf{NGC 6860}: Suggested as double-barred by \citet{marquez99},
based on features in their near-IR ellipse fits and color maps. 
Because the ellipse-fit features ($a < 2\arcsec$) are only slightly
larger than their seeing (FWHM $\approx 0.9\arcsec$), this is
unfortunately not a clear detection; higher-resolution images are
clearly needed.  (The available \textit{HST} images are off-center or 
too low in signal to be useful.)

\textbf{NGC 6907}: Suggested as double-barred by \citet{elmegreen96}.
Examination of the publically available $B$-band and $H$-band images
from the OSU Bright Spiral Galaxy Survey \citep{eskridge02} shows a
clear elliptical feature perfectly aligned with the outer disk, with
apparently somewhat higher ellipticity (so this is not an inner disk).
Unfortunately, available images do not have high enough resolution to
rule out the possiblity of, e.g., a nuclear ring.  The (outer?)  bar is
relatively short; ellipse-fit measurements are confused due to the
strong spiral arms trailing off from the bar ends.

\textbf{NGC 7742}: Suggested as double-barred by \citet{laine02},
though \citet{w95} found no bars in their optical study.  Inspection
and unsharp masking of a NICMOS2 F160W image yields some evidence for a
weak nuclear bar, matching the inner bar of Laine et al.  Their outer
bar, unfortunately, is difficult to confirm: it is either
extraordinarily weak and round, or else an isophotal side effect of the
bright, star-forming ring and its associated dust lanes.

\textbf{IC 454}: Suggested as double-barred by \citet{marquez99},
based on features in their near-IR ellipse fits.  This is a promising 
candidate, but since the size of the apparent inner bar ($a = 
3\arcsec$) is only a little larger than their resolution (seeing = 
1.2--1.3\arcsec), it should probably be confirmed with higher 
resolution observations.

\textbf{IC 1816}: Suggested as double-barred by \citet{marquez99},
though they were uncertain about the inner bar's existence. 
Examination, ellipse fits, and unsharp masking of a WFPC2 F606W image
indicates that the inner ellipticity peak is most likely due to a
chaotic, possibly star-forming nuclear ring and associated dust lanes;
high-resolution near-IR imaging needed to clearly determine if there
is or is not an inner bar \textit{inside} the nuclear ring.
  
\textbf{IC 2510}: Suggested as a possible double bar by
\citet{marquez99}; the inner bar detection was based on only one
method, so they considered it ``doubtful.''  The WFPC2 F606W image
shows a chaotic morphology dominated by dust; higher-resolution
near-IR images are probably needed to determine if there is an inner
bar in this galaxy.
  
\textbf{Mrk 1066}: Suggested as double-barred by \citet{laine02}; also
listed by \citet{moiseev01}, on the basis of optical images by
\citet{afanasiev98}.  Inspection of NICMOS2 F160W and F205W images
shows an elongated central region, matching the inner bar of Laine et
al.; unfortunately, strong isophote twisting, dust lanes, and
IR-bright spiral structure make it difficult to be certain this is a
bar, rather than an inclined or distorted pair of spiral arms.

\textbf{UGC 3223}: Suggested as double-barred by \citet{marquez99},
though they were uncertain about the inner bar's existence; their
ellipse fits just barely suggest something might be there.  Nothing
barlike is apparent in a WFPC2 F606W image, though there are strong
dust lanes.  Higher-resolution infrared images are needed.

\textbf{ESO 323-77}: \citet{greusard00} listed this as a ``bar +
triaxial bulge'' object; the orientation of the inner ellipticity peak
suggests a possible inner disk (PA = 156\degr, versus 155\degr{}
for outer disk from RC3).  Inspection of WFPC2 F606W images shows
tightly wrapped spiral arms in the nuclear region, with some bright
knots possibly indicating star formation.  High-resolution near-IR
imaging is needed to determine if the near-IR isophote twists seen by
Greusard et al.\ are solely due to the spiral arms.

\textbf{ESO 437-67}: Suggested by \citet{jungwiert97}.  However, the
proposed secondary bar is too small to be securely identified; higher
resolution images needed, as Jungwiert et al.\ note.

\textbf{ESO 508-78}: Suggested by \citep{bc91,bc93}.  Inspection of 
images kindly provided by Ron Buta shows a possible elliptical feature 
in the galaxy center, but the resolution is not good enough to clearly 
determine whether it is an inner bar or not.

\textbf{Milky Way}: \citet{alard01} recently presented evidence for an
asymmetric planar structure in the central 2--3\degr{} of the
galactic center, seen in near-IR star counts from 2MASS \textit{after}
subtracting the contribution of the known, large-scale bar.  The
asymmetry suggests a different orientation than that of the
large-scale bar, which might indicate a small bar seen at an an
intermediate position angle.  However, \citet{vanloon03} argue, from
an analysis of ISOGAL and DENIS data, that the inner $\sim 1$ kpc of
the Galaxy is axisymmetric.

\section{Misclassified Single-Bar (or Unbarred) Galaxies}
\label{sec:false}

The following list contains galaxies which have been suggested as
being double-barred or inner-disk galaxies, but which a careful
analysis (often using archival \textit{HST} images) indicates are
probably \textit{not}.  In many cases, there \textit{is} a distinct
feature inside a large-scale bar which masquerades as an inner bar,
but it is a nuclear ring or nuclear spiral -- or simply strong dust
lanes -- instead.

\textbf{NGC 1300}: Suggested as a possible double bar by
\citet{moiseev01}, on the basis of optical plates in
\citet{baumgart86}.  Using ground-based and NICMOS near-IR images,
\citet{perez-r00} found a nuclear ring, but no evidence for an inner
bar.

\textbf{NGC 1365}: Suggested as double-barred by \citet{jungwiert97}
and \citet{laine02}, using ground-based and NICMOS images,
respectively.  Inspection of the NICMOS images shows that the
ellipticity peak is due to a nuclear ring, with no evidence for a bar
inside; a similar conclusion was reached by \citet{emsellem01} based
on stellar kinematics from long-slit spectroscopy.

\textbf{NGC 1512}: Listed by \citet{moiseev01}, on the basis of weak
ellipse-fit features noted by \citet{jungwiert97} in their near-IR
images (which they interpreted as evidence for a triaxial bulge). 
Examination of NICMOS2 archival images (F160W and F187W) shows a
narrow, star-forming nuclear ring, with no evidence for any bar
inside.  The slight PA and ellipticity variations seen by Jungwiert et
al.\ are almost certainly due to the nuclear ring.

\textbf{NGC 1530}: Suggested as double-barred by \citet{laine02} on
the basis of ellipse fits to NICMOS images.  However,
\citet{perez-r00}, using ground-based near-IR images, argued that a
nuclear spiral is main inner feature.  Inspection of the NICMOS2 F160W
image and unsharp masks strongly suggests a nuclear spiral rather than
a bar \citep[see also][]{martini03-apjs}.  There is some suggestion
from both the F160W image and from the WFPC2 F606W images that the
innermost spiral may be lopsided, with one arm brighter than the
other.

\textbf{NGC 1566}: Listed by \citet{moiseev01}, based on the
suggestion of a triaxial bulge by \citet{baumgart86}.  Inspection of
F606W and F814W WFPC2 images shows strong dust lanes leading to a
nuclear spiral in the inner $r < 3\arcsec$ region.  The $K$-band image
of \citet{mrk97} shows no inner ellipticity peak and only a slight
twist in the position angle, most likely due to the nuclear spiral.

\textbf{NGC 1667}: Suggested as double-barred by \citet{laine02}. 
Inspection of the $K$-band image of \citet{mrk97}, the NICMOS2 F160W
image, and the WFPC2 F606W image all indicate that ``outer bar'' is
really a pair of spiral arms (as noted by Mulchaey et al.); see
Fig.~\ref{fig:n1667}.  In this case, it is the ``inner'' bar which is
evidently the genuine (and only) bar in this galaxy; isophote twists
interior to \textit{it} are due to strong dust lanes.
  
\textbf{NGC 1672}: Listed by \citet{moiseev01}, based on the
suggestion of a triaxial bulge by \citet{baumgart86}.  Inspection of
F814W WFPC2 and F160W NICMOS2 images show a broad, star-forming
nuclear ring, with a nuclear spiral inside \citep[see
also][]{martini03-apjs}.  Although the latter is dusty, the dust is
confined to the eastern side; there is no sign of an inner bar.

\textbf{NGC 2273}: Suggested as double-barred by \citet{mrk97}. 
Optical and near-IR \textit{HST} images show inner bar is nuclear ring
with two-armed spiral inside \citep[see Fig.~\ref{fig:nucrings1} and
][]{erwin-sparke03,martini03-apjs}.

\textbf{NGC 2712}: Suggested as double-barred by \citet{marquez99},
based on their near-IR images.  In this case, inspection of
unpublished, higher-resolution $J$ and $K$-band images from the
William Herschel Telescope (Erwin, Vega Belt\'an, \& Beckman, in prep)
shows that the inner elliptical feature identified by M\'arquez et
al.\ is a somewhat irregular nuclear ring; there is no evidence for a
bar inside this ring.

\textbf{NGC 3359}: \citet{sempere99} and \citet{rozas00} suggested
that this galaxy might have two decoupled bars, based on attempts to
match hydrodynamic simulations to \hi{} and \ha{} morphology, and on
evidence of isophote twisting inside the ``main'' bar in an $I$-band
image.  However, analysis of a deep $I$-band image from the BARS
Project \citep{lourenso01} and of the $K$-band image of \citet{mh01}
shows evidence for only one (stellar) bar, with a radius of $\sim
15$--20\arcsec.  The isophote twisting in the $I$ band is absent at
$K$, and is almost certainly due to dust rather than to an inner bar. 
(This does not rule out multiple pattern speeds in the galaxy,
however: the two pattern speeds reported by Rozas \& Sempere could be
those of the bar and the outer spirals, respectively.)

\textbf{NGC 3786}: Listed in \citet{moiseev01}, on the basis of
optical images in \citet{afanasiev98}.  However, \citet{moiseev02}
reported that the ``outer bar'' is really spiral arms, and there is no
sign of any large-scale bar in the WFPC2 F606W image.

\textbf{NGC 4371}: \citet{w95} suggested this galaxy might be
\textit{triply} barred \citep[it is listed as such by][]{moiseev01},
but pointed out that high inclination and alignment of inner two
``bars'' with outer disk could mean that they were projected
axisymmetric structures; \citet{kormendy79} noted a possible secondary
bar in this galaxy, which corresponds to the middle of W95's three
ellipticity peaks.  \citet{erwin99} showed that the middle ellipticity
peak was due to a bright, stellar ring; the apparent inner peak was
produced by the effects of superimposing a ring on a rounder bulge
profile, accompanied by circularizing of the inner isophotes by seeing
\citep[see Fig.~\ref{fig:nucrings2} and ][]{erwin01-lapalma}.

\textbf{NGC 4593}: Listed by \citet{moiseev01}, based on the suggestion
of a triaxial bulge by \citet{w95}.  The latter authors noted the
presence of dust and a blue nuclear ring with $r \sim 2\arcsec$.
Inspection of a NICMOS2 F160W confirms that the central region of this
galaxy is dominated by spiral dust lanes surrounding a nuclear ring or
pseudo-ring; there is no evidence for an inner bar.

\textbf{NGC 4643}: Listed as an ``inner-disk'' galaxy by
\citet{erwin-sparke03}; higher-resolution $I$-band images (courtesy
Johan Knapen) and unpublished $H$-band images from the William
Herschel Telescope appear to confirm the alternate hypothesis of Erwin
\& Sparke that this is a stellar nuclear ring instead of an inner
disk.

\textbf{NGC 5033}: Suggested as a \textit{triple} bar system by
\citet{laine02}.  \citet{martini01} did not find any bars in their
analysis of the same images, and \citet{laurikainen02} found no sign
of a bar in their tangential-force analysis of this galaxy using 2MASS
images.  Inspection of \textit{HST} NICMOS1 and NICMOS2 images (and a
WFPC2 F606W image) shows extremely strong spiral dust lanes in a
highly inclined galaxy, with small, luminous bulge in center
\citep[see also][]{martini03-apjs}.  The $K$-band image of
\citet{peletier99} \textit{does} suggest that the outermost bar of
Laine et al.\ might be real, but the inner ellipticity peaks are
almost certainly due to dust.

\textbf{NGC 5383}: Suggested as a potential double bar by
\citet{regan97} on basis of ground-based $K$-band images, though they
noted possibility of ``projection effects.''  Also reported as
double-barred by \citet{laine02}.  However, \citet{sheth00}, using
NICMOS2 images, showed that the inner feature is a nuclear spiral.

\textbf{NGC 6221}: Suggested as double-barred by \citet{greusard00},
based on ellipse fits to their ground-based near-IR images; the
orientation is the same as the outer disk, so this would be an inner
disk candidate.  However, NICMOS2 F160W images indicate that a nuclear
spiral is most likely the cause of the inner ellipticity peak.

\textbf{NGC 6300}: Both \citet{mrk97} and \citet{laine02} have
suggested that this galaxy is double-barred (though with disagreements
about the size of the inner bar).  Inspection of the NIMCOS2 F160W
image shows that the inner isophotes are indeed elongated and twisted;
however, these are distortions caused by a strong dust lane (clearly
visible in WFPC2 images).

\textbf{NGC 6951}: Suggested as double-barred by \citet{w95} and
\citet{mrk97}.  However, \citet{friedli96a} noted that it might be
nuclear ring instead, and NICMOS2 F110W and F160W images confirm the
latter interpretation
\citep[e.g.,][]{perez00,perez-r00,martini03-apjs}.

\textbf{NGC 7743}: \citet{regan99} suggested this as double-barred
based on the appearance of straight dust lanes.  However,
\citet{shlosman02} and \citet{witold02} have used hydrodynamical
simulations to argue that inner bars are unlikely to have straight
dust lanes, so identifications based solely on the shape of dust lanes
are suspect.  \citet{erwin-sparke03} argued that the ``straight'' dust
lanes were really part of a nuclear spiral.

\textbf{Mrk 471}: Suggested as double-barred by \citet{martini99} and
\citet{martini01}, based on NICMOS1 F160W images.  Inspection of both
those images and the F110W images indicates that the isophote
distortions (and the resulting ellipse fits) are due to strong,
asymmetric dust extinction; there are no other indications of a
barlike structure in the central regions.

\textbf{IC 184}: Suggested as double-barred by \citet{marquez99}. 
Inspection of the WFPC2 F606W image shows the ``inner bar'' quite
clearly as the galaxy's \textit{only} bar; the ``outer bar'' is
resolved into a series of bright spiral arms (thus, this galaxy is
similar to NGC~1667, above).  There is no sign of any bar inside the
``inner'' bar; there is an inner ellipticity peak in the WFPC2 ellipse
fits, but this is apparently due to dust lanes (high-res near-IR
imaging may be needed to clearly rule out a true inner bar).

\textbf{UGC 1395}: Suggested as double-barred by \citet{marquez99}, on
the basis of unsharp masking.  There is no sign of any structure with
the size and position angle they suggest in the F606W WFPC2 image, or
in the NICMOS1 F160W images; strong dust lanes in the bar may be
confusing things.  \citet{martini99} did not report finding a bar, and
there is no evidence for a bar in published near-IR images of
\citet{peletier99}.

\begin{acknowledgements}
I would like to thank Marc Balcells, Linda Sparke, Andrew Cardwell,
and John Beckman for interesting discussions and helpful comments on
early drafts of the work.  I am also grateful to several people who
provided me with images of various candidate double-barred galaxies,
including Ron Buta, Fran\c{c}oise Combes, Chris Conselice and Jay
Gallagher, Johan Knapen, Alfonso L\'opez Aguerri and Enrico Maria Corsini,
Oak-Kyoung Park, and Paul Schechter.

This work made use of data from the Ohio State University Bright
Spiral Galaxy Survey, which was funded by grants AST-9217716 and
AST-9617006 from the United States National Science Foundation, with
additional support from Ohio State University.  It also made use of
images from the Barred and Ringed Spirals (BARS) database, for which
time was awarded by the Comit\'e Cient\'{\i}fico Internacional of the
Canary Islands Observatories.  Based on observations made with the
NASA/ESA Hubble Space Telescope, obtained from the data archive at the
Space Telescope Science Institute.  STScI is operated by the
Association of Universities for Research in Astronomy, Inc.  under
NASA contract NAS 5-26555.

Finally, this research made use of the Lyon-Meudon Extragalactic
Database (LEDA; part of HyperLeda at http://leda.univ-lyon1.fr/), and
the NASA/IPAC Extragalactic Database (NED), which is operated by the
Jet Propulsion Laboratory, California Institute of Technology, under
contract with the National Aeronautics and Space Administration.

\end{acknowledgements}



\begin{table*}
\begin{minipage}{126mm}
    \caption{Measured Values for Bars and Inner Disks}
    \label{tab:measured}
    \begin{tabular}{lrrrrrrrr}
\hline
Name & Outer Disk PA & Bar/Disk PA &  \amax   & \amin   & \aten   & \lbar 
  & \emax & NR? \\
     & \degr{}       & \degr{}     &  \arcsec & \arcsec & \arcsec & \arcsec
  &  & \\
(1)  & (2)           & (3)         & (4)      & (5)     & (6)     & (7)
  & (8)   & (9) \\
\hline
\multicolumn{9}{c}{Double-Barred Galaxies} \\
\hline
NGC 357 &   20   &  120 &   21 &   28 &   27 &   27 &  0.44 &   \\*
         &        &   45 & 3.1 & 5.7 & 4.4 & 4.4 &  0.16 &   \\[3mm]
 NGC 718 &    5   &  152 &   20 &   33 &   30 &   30 &  0.23 & Y \\*
         &        &   15 & 1.6 & 4.0 & 3.6 & 3.3 &  0.19 &   \\[3mm]
NGC 1068 &   98   &   12 &   54 &   75 &   89 &   75 &  0.24 & Y \\*
         &        &   47 &   15 &   27 & \ldots  &   17 &  0.45 &   \\[3mm]
NGC 1097 &  134   &  147 &   88 &  120 & \ldots  &  107 &  0.67 & Y \\*
         &        &   30 & 7.5 & 8.1 & 7.7 & 7.7 &  0.46 &   \\[3mm]
NGC 1241 &  145   &  110 &   18 &   24 &   23 &   18 &  0.60 & Y \\*
         &        &    0 & 1.5 & \ldots  & 2.1 & 1.8 &  0.31 &   \\[3mm]
NGC 1291 & \ldots  &  171 &   89 &  130 &  140 &  130 &  0.39 &   \\*
         &        &   15 &   18 &   26 &   24 &   24 &  0.24 &   \\[3mm]
NGC 1317 &   78   &  150 &   41 &   59 &   58 &   58 &  0.27 & Y \\*
         &        &   56 & 6.3 &   11 & 9.7 & 6.4 &  0.44 &   \\[3mm]
NGC 1433 &   21   &   97 &   74 &  110 &  150 &   90 &  0.70 & Y \\*
         &        &   32 & 6.2 &   13 &   12 &   12 &  0.38 &   \\[3mm]
NGC 1543 & \ldots  &   92 &   63 &   95 & \ldots  &   95 &  0.49 & Y \\*
         &        &   26 & 7.9 &   12 &   12 &   11 &  0.29 &   \\[3mm]
NGC 1808 &  133   &  144 &   80 &  114 & \ldots  &  114 &  0.66 & Y \\*
         &        &  158 & 3.3 & 5.9 & 4.8 & 4.8 &  0.53 &   \\[3mm]
NGC 2217 &    5   &  136 &   37 & \ldots  & \ldots  &   50 &  0.47 &   \\*
         &        &  112 & 7.8 &   11 &   14 &   11 &  0.19 &   \\[3mm]
NGC 2642 &  140   &  115 &   25 & \ldots  &   29 &   26 &  0.64 & Y \\*
         &        &  145 & 1.5 & 2.7 & 2.7 & 2.0 &  0.30 &   \\[3mm]
NGC 2646 & \ldots  &   82 &   16 &   21 &   21 &   21 &  0.48 &   \\*
         &        &    8 & 2.2 & 4.6 & 3.5 & 3.5 &  0.25 &   \\[3mm]
NGC 2681 &  140 &   30 &  50 &  75 &   60 &   60 &  0.23 & Y \\*
         &      &   73 &  18 &  23 &   19 &   19 &  0.33 & \\*
         &      &   20 & 1.7 & 3.9 &  3.3 &  3.3 &  0.26 & \\[3mm]
NGC 2859 &   90   &  162 &   34 &   52 &   43 &   43 &  0.40 & Y \\*
         &        &   62 & 4.1 &   11 & 6.2 & 6.2 &  0.31 &   \\[3mm]
NGC 2950 &  120   &  162 &   24 &   41 &   31 &   31 &  0.43 & Y \\*
         &        &   85 & 3.2 & 6.3 & 3.9 & 3.9 &  0.33 &   \\[3mm]
NGC 2962 &   10   &  168 &   29 &   43 & \ldots  &   43 &  0.30 &   \\*
         &        &   93 & 3.5 & \ldots  & 4.2 & 4.2 &  0.03 &   \\[3mm]
\hline
\end{tabular}
\medskip

Observed position angles, sizes, and ellipticities for bars and inner
disks of double-barred and inner-disk galaxies.  For each galaxy, the
first line contains values for the outer bar and the second line
contains values for the inner bar (of double-barred galaxies) or inner
disk.  Col.\ (1): Galaxy name.  Col.\ (2): Position angle of the outer
disk, from Table~\ref{tab:galaxies-db} or \ref{tab:galaxies-id}. 
Col.\ (3): Bar/disk position angle.  Col.\ (4): Bar/disk semi-major
axis of maximum isophotal ellipticity \amax{} (lower limit on bar
size).  Col.\ (5): Semi-major axis of minimum ellipticity \amin{}
outside \amax.  Col.\ (6): Semi-major axis where $|$PA $-$ bar PA$| >
10\degr$ outside \amax.  Col.\ (7): \lbar{} (upper limit on bar size
-- usually the minimum of \amin{} and \aten{}; see
Sect.~\ref{sec:notes} for sources of alternate values).  Col.\ (8):
Maximum isophotal ellipticity ($\epsilon = 1 - b/a$) of the bar or
inner disk.  Col.\ (9): Nuclear ring present?  Note that NGC~3368,
3945, 4785, and 7187 have both inner bars \textit{and} inner disks,
and are thus included in each section.

\end{minipage}
\end{table*}

\setcounter{table}{2}
\begin{table*}
\begin{minipage}{126mm}
    \caption{Continued}
    \begin{tabular}{lrrrrrrrr}
\hline
Name & Outer Disk PA & Bar/Disk PA &  \amax   & \amin   & \aten   & \lbar 
  & \emax & NR? \\
     & \degr{}       & \degr{}     &  \arcsec & \arcsec & \arcsec & \arcsec
  &  & \\
(1)  & (2)           & (3)         & (4)      & (5)     & (6)     & (7)
  & (8)   & (9) \\
\hline
NGC 3081 &   97   &   69 &   33 &   41 & \ldots  &   35 &  0.65 & Y \\*
         &        &  122 & 5.7 & 8.6 & 7.1 & 5.8 &  0.50 &   \\[3mm]
NGC 3275 &  122   &  120 &   28 &   36 &   48 &   36 &  0.58 & Y \\*
         &        &  172 & 2.1 & 3.5 & 2.7 & 2.6 &  0.31 &   \\[3mm]
NGC 3358 &  125   &   98 &   16 &   21 &   24 &   21 &  0.42 &   \\*
         &        &  136 & 4.9 & 8.3 &   11 & 8.3 &  0.32 &   \\[3mm]
NGC 3368 &  172   &  115 &   61 &   80 &   75 &   75 &  0.40 & Y \\*
         &        &  129 & 3.4 & 6.8 & 5.9 & 5.0 &  0.35 &   \\[3mm]
NGC 3393 & \ldots  &  157 &   13 &   16 &   23 &   16 &  0.44 &   \\*
         &        &  146 & 1.9 & 3.5 & 3.1 & 3.1 &  0.20 &   \\[3mm]
NGC 3941 &   10   &  166 &   21 &   36 &   32 &   32 &  0.47 &   \\*
         &        &   85 & 3.2 & 4.7 & 4.4 & 4.4 &  0.21 &   \\[3mm]
NGC 3945 &  158   &   72 &   32 &   41 &   39 &   39 &  0.29 & Y \\*
         &        &   90 & 2.6 & 3.0 & 3.0 & 3.0 &  0.11 &   \\[3mm]
NGC 4303 &  138   &  179 &   29 &   50 &   48 &   34 &  0.50 & Y \\*
         &        &   40 & 1.8 & 2.8 & 2.5 & 2.5 &  0.29 &   \\[3mm]
NGC 4314 &   65   &  146 &   67 &   90 &  111 &   80 &  0.64 & Y \\*
         &        &  136 & 4.5 & 5.8 & 5.6 & 5.6 &  0.23 &   \\[3mm]
NGC 4321 &  153   &  107 &   55 &   61 &   80 &   58 &  0.54 & Y \\*
         &        &  110 & 8.2 &   15 &   10 &   10 &  0.62 &   \\[3mm]
NGC 4340 &   95   &   31 &   39 &   51 &   48 &   48 &  0.39 & Y \\*
         &        &   25 & 3.4 & 5.1 & 4.7 & 4.5 &  0.11 &   \\[3mm]
NGC 4503 &   12   &    6 &   23 &   27 & \ldots  &   27 &  0.48 &   \\*
         &        &   45 & 2.9 & 4.8 & 6.0 & 4.8 &  0.25 &   \\[3mm]
NGC 4725 &   40   &   50 &  118 &  130 &  170 &  125 &  0.67 &   \\*
         &        &  141 & 5.6 & 7.3 & 6.8 & 6.8 &  0.20 &   \\[3mm]
NGC 4736 &  113   &   90 &  125 &  170 & \ldots  &  170 &  0.23 & Y \\*
         &        &   25 &   11 &   21 &   20 &   20 &  0.22 &   \\[3mm]
NGC 4785 &   81   &   66 &   11 &   13 &   16 &   12 &  0.57 & Y \\*
         &        &  127 & 0.9 & 1.1 & 1.2 & 1.1 &  0.21 &   \\[3mm]
NGC 4984 &   15   &   95 &   30 & \ldots  &   38 &   32 &  0.30 & Y \\*
         &        &   64 & 4.0 & 5.0 & 4.9 & 4.9 &  0.23 &   \\[3mm]
NGC 5365 &    4   &  108 &   25 &   33 &   31 &   31 &  0.25 & Y \\*
         &        &   35 & 3.7 &   16 & 5.3 & 5.3 &  0.28 &   \\[3mm]
NGC 5728 &    2   &   33 &   56 &   71 & \ldots  &   71 &  0.71 & Y \\*
         &        &   86 & 1.8 & 4.2 & 3.7 & 3.7 &  0.49 &   \\[3mm]
NGC 5850 &  163   &  116 &   63 &   84 & \ldots  &   75 &  0.68 & Y \\*
         &        &   50 & 5.9 & 9.2 & 7.6 & 7.1 &  0.30 &   \\[3mm]
NGC 6654 &    0   &   17 &   26 &   47 &   38 &   38 &  0.51 &   \\*
         &        &  135 & 2.7 & 4.4 & 4.2 & 4.2 &  0.15 &   \\[3mm]
NGC 6684 &   36   &  150 &   26 &   33 &   29 &   28 &  0.25 &   \\*
         &        &   68 & 2.9 & 6.2 & 4.1 & 4.1 &  0.35 &   \\[3mm]
NGC 6782 &   45   &  178 &   25 &   30 &   33 &   30 &  0.54 & Y \\*
         &        &  147 & 3.4 & 4.5 & 4.3 & 3.8 &  0.47 &   \\[3mm]
\hline
\end{tabular}
\end{minipage}
\end{table*}

\setcounter{table}{2}
\begin{table*}
\begin{minipage}{126mm}
    \caption{Continued}
    \begin{tabular}{lrrrrrrrr}
\hline
Name & Outer Disk PA & Bar/Disk PA &  \amax   & \amin   & \aten   & \lbar 
  & \emax & NR? \\
     & \degr{}       & \degr{}     &  \arcsec & \arcsec & \arcsec & \arcsec
  &  & \\
(1)  & (2)           & (3)         & (4)      & (5)     & (6)     & (7)
  & (8)   & (9) \\
\hline
NGC 7098 &   75   &   46 &   38 &   53 &   56 &   53 &  0.57 &   \\*
         &        &   71 & 9.1 &   12 &   18 &   12 &  0.32 &   \\[3mm]
NGC 7187 &  134   &   66 &   19 &   28 &   22 &   19 &  0.37 & Y \\*
         &        &   54 & 2.5 & 3.8 & 3.5 & 3.5 &  0.19 &   \\[3mm]
NGC 7280 &   72   &   55 &   11 &   29 &   27 &   27 &  0.40 &   \\*
         &        &  115 & 1.2 & 2.2 & 1.6 & 1.6 &  0.30 &   \\[3mm]
NGC 7716 &   35   &   25 &   22 &   28 &   25 &   25 &  0.35 &   \\*
         &        &   57 & 3.1 & 5.4 & 5.6 & 5.4 &  0.38 &   \\[3mm]
 Mrk 573 & \ldots  &    0 & 9.0 &   12 & \ldots  &   12 &  0.33 & Y \\*
         &        &   83 & 1.2 & 3.1 & \ldots  & 2.1 &  0.31 &   \\[3mm]
 UGC 524 &  120   &  143 & 8.5 &   11 &   12 &   11 &  0.54 & Y \\*
         &        &  167 & 1.0 & 1.1 & 1.3 & 1.1 &  0.43 &   \\[3mm]
ESO 215-G031 &  130   &  147 &   38 & \ldots  & \ldots  &   47 &  0.63 & Y \\*
         &        &  153 & 5.3 & 9.5 & 6.2 & 6.2 &  0.48 &   \\[3mm]
ESO 320-G030 &  121   &  142 &   23 & \ldots  &   41 &   37 &  0.64 &   \\*
         &        &  107 & 3.9 & 5.2 & 4.5 & 4.5 &  0.32 &   \\[3mm]
ESO 443-G039 &   14   &   27 &   11 &   19 &   25 &   19 &  0.51 &   \\*
         &        &   44 & 1.8 & 2.4 & 2.7 & 2.4 &  0.24 &   \\[3mm]
ESO 447-G030 &   35   &  133 &   13 &   15 &   14 &   14 &  0.17 & Y \\*
         &        &  177 & 2.8 & 9.7 & 3.6 & 3.6 &  0.36 &   \\[3mm]
IRAS 03565+2139 & \ldots  &    4 &   13 &   18 &   20 &   15 &  0.60 & Y \\*
         &        &  124 & 1.6 & 3.3 & 3.0 & 1.9 &  0.33 &   \\[3mm]
\hline
\multicolumn{4}{c}{Inner-Disk Galaxies} \\
\hline
 NGC 151 &   75   &  152 &   18 &   23 &   21 &   21 &  0.44 & Y \\*
         &        &   74 & 4.3 & 7.1 & 6.1 & 6.1 &  0.32 &   \\[3mm]
 NGC 470 &  155   &   18 &   21 &   30 &   26 &   26 &  0.55 &   \\*
         &        &  152 & 2.7 & 7.4 &   11 & 7.4 &  0.46 &   \\[3mm]
NGC 1398 &   96   &    9 &   38 &   46 &   44 &   44 &  0.36 &   \\*
         &        &   96 & 2.9 & 8.2 & 7.9 & 7.9 &  0.18 &   \\[3mm]
NGC 2787 &  109   &  160 &   29 &   36 &   36 &   36 &  0.34 &   \\*
         &        &  113 &   18 &   22 &   21 &   21 &  0.34 &   \\[3mm]
NGC 2880 &  144   &   82 & 8.0 & 9.0 &   10 & 9.0 &  0.20 &   \\*
         &        &  138 & 3.0 & 5.6 & 4.5 & 4.5 &  0.22 &   \\[3mm]
NGC 3266 &   85   &    8 &   10 &   14 &   13 &   13 &  0.29 & Y \\*
         &        &   90 & 0.8 & 2.4 & 1.8 & 1.8 &  0.12 &   \\[3mm]
NGC 3368 &  172   &  115 &   61 &   80 &   75 &   75 &  0.40 & Y \\*
         &        &  162 &   21 &   35 &   30 &   30 &  0.30 &   \\[3mm]
NGC 3384 &   50   &  132 &   15 & \ldots  &   17 &   17 &  0.05 &   \\*
         &        &   46 & 2.7 &   15 &   11 &   11 &  0.40 &   \\[3mm]
NGC 3412 &  153   &  100 &   15 &   21 &   21 &   21 &  0.26 &   \\*
         &        &  154 & 1.0 & 8.3 & 6.1 & 6.1 &  0.33 &   \\[3mm]
NGC 3945 &  158   &   72 &   32 &   41 &   39 &   39 &  0.29 & Y \\*
         &        &  158 &   10 &   19 &   18 &   18 &  0.36 &   \\[3mm]
NGC 4143 &  144   &  163 &   17 & \ldots  &   28 &   28 &  0.38 &   \\*
         &        &  142 & 4.2 & 6.2 & \ldots  & 6.2 &  0.25 &   \\[3mm]
\hline
\end{tabular}
\end{minipage}
\end{table*}

\setcounter{table}{2}
\begin{table*}
\begin{minipage}{126mm}
    \caption{Continued}
    \begin{tabular}{lrrrrrrrr}
\hline
Name & Outer Disk PA & Bar/Disk PA &  \amax   & \amin   & \aten   & \lbar 
  & \emax & NR? \\
     & \degr{}       & \degr{}     &  \arcsec & \arcsec & \arcsec & \arcsec
  &  & \\
(1)  & (2)           & (3)         & (4)      & (5)     & (6)     & (7)
  & (8)   & (9) \\
\hline
NGC 4262 &  153   &   18 &   13 &   20 &   16 &   16 &  0.34 &   \\*
         &        &  155 & 3.5 & 4.4 & 4.7 & 4.4 &  0.10 &   \\[3mm]
NGC 4386 &  140   &  134 &   25 &   36 & \ldots  &   36 &  0.52 &   \\*
         &        &  141 & 2.4 & 3.2 & \ldots  & 3.2 &  0.28 &   \\[3mm]
NGC 4612 &  143   &   83 &   17 &   24 &   20 &   20 &  0.22 &   \\*
         &        &  144 & 3.0 & 5.7 & 5.5 & 5.5 &  0.21 &   \\[3mm]
NGC 4754 &   23   &  142 &   23 &   27 &   30 &   27 &  0.23 &   \\*
         &        &   22 & 6.7 & 9.4 &   12 & 9.4 &  0.23 &   \\[3mm]
NGC 4785 &   81   &   66 &   11 &   13 &   16 &   12 &  0.57 & Y \\*
         &        &   83 & 5.2 & 6.7 & 8.4 & 6.7 &  0.41 &   \\[3mm]
NGC 7007 &    2   &  116 & 6.2 & 8.1 & 7.6 & 7.6 &  0.26 &   \\*
         &        &    0 & 2.3 & 3.7 & 2.9 & 2.9 &  0.20 &   \\[3mm]
NGC 7187 &  134   &   66 &   19 &   28 &   22 &   19 &  0.37 & Y \\*
         &        &  134 & 7.2 & 9.4 & 8.6 & 8.6 &  0.12 &   \\[3mm]
UGC 6062 &   25   &  159 &   12 &   17 &   15 &   15 &  0.39 &   \\*
         &        &   27 & 1.9 & 2.8 & 3.4 & 2.8 &  0.39 &   \\[3mm]
ESO 378-G020 &   34   &   81 & 6.9 & 8.3 & 8.8 & 8.3 &  0.37 &   \\*
         &        &   39 & 0.8 & 1.3 & \ldots  & 1.3 &  0.42 &   \\[3mm]
ESO 443-G017 &   23   &  160 & 9.6 &   15 &   12 &   12 &  0.48 &   \\*
         &        &   31 & 2.3 & 5.5 & 4.2 & 4.2 &  0.28 &   \\[3mm]
\hline
\end{tabular}
\end{minipage}
\end{table*}


\begin{table*}
\begin{minipage}{126mm}
\caption{Deprojected Values for Bars and Inner Disks}
\label{tab:deproj}
\begin{tabular}{lrrr|lrrr}
\hline
Name & $\Delta$PA & Bar/Disk Size & Bar/Disk Size & 
Name & $\Delta$PA & Bar/Disk Size & Bar/Disk Size \\
     & \degr{}    & kpc           & f($R_{25})$   &
     & \degr{}    & kpc           & f($R_{25})$ \\
(1)  & (2)        & (3)           & (4) &
(1)  & (2)        & (3)           & (4) \\
\hline
\multicolumn{8}{c}{Double-Barred Galaxies} \\
\hline
 & & & & & & &\\
 NGC 357 &    67 &  4.01,  5.15 &  0.36,  0.47  & NGC 3081 &    62 &  5.02,  5.32 &  0.55,  0.58\\*
         &       &  0.50,  0.71 &  0.05,  0.06  &          &       &  0.86,  0.87 &  0.09,  0.10\\[3mm]
 NGC 718 &    48 &  2.30,  3.45 &  0.30,  0.44  & NGC 3275 &    60 &  5.50,  7.07 &  0.33,  0.42\\*
         &       &  0.18,  0.36 &  0.02,  0.05  &          &       &  0.50,  0.62 &  0.03,  0.04\\[3mm]
NGC 1068 &    28 &  5.02,  6.97 &  0.33,  0.46  & NGC 3358 & $-$49 &  3.13,  4.10 &  0.17,  0.23\\*
         &       &  1.28,  1.45 &  0.08,  0.10  &          &       &  0.90,  1.52 &  0.05,  0.09\\[3mm]
NGC 1097 & $-$61 &  6.26,  7.62 &  0.32,  0.39  & NGC 3368 & $-$11 &  4.39,  5.40 &  0.38,  0.47\\*
         &       &  0.74,  0.76 &  0.04,  0.04  &          &       &  0.22,  0.33 &  0.02,  0.03\\[3mm]
NGC 1241 & $-$78 &  5.88,  5.88 &  0.28,  0.28  & NGC 3393 & $-$11 &  3.09,  3.80 &  0.20,  0.25\\*
         &       &  0.49,  0.59 &  0.02,  0.03  &          &       &  0.46,  0.75 &  0.03,  0.05\\[3mm]
NGC 1291 &    24 &  3.41,  4.98 &  0.30,  0.44  & NGC 3941 & $-$64 &  1.39,  2.12 &  0.23,  0.34\\*
         &       &  0.69,  0.92 &  0.06,  0.08  &          &       &  0.29,  0.41 &  0.05,  0.07\\[3mm]
NGC 1317 & $-$80 &  4.38,  6.19 &  0.56,  0.80  & NGC 3945 & $-$11 &  4.65,  5.67 &  0.32,  0.38\\*
         &       &  0.60,  0.61 &  0.08,  0.08  &          &       &  0.36,  0.42 &  0.02,  0.03\\[3mm]
NGC 1433 &    65 &  4.58,  5.57 &  0.45,  0.55  & NGC 4303 & $-$41 &  2.25,  2.64 &  0.16,  0.18\\*
         &       &  0.33,  0.63 &  0.03,  0.06  &          &       &  0.15,  0.20 &  0.01,  0.01\\[3mm]
NGC 1543 &  (66) &  6.50,  9.80 &  0.46,  0.69  & NGC 4314 & $-$10 &  4.29,  5.12 &  0.59,  0.70\\*
         &       &  0.78,  1.08 &  0.05,  0.08  &          &       &  0.29,  0.36 &  0.04,  0.05\\[3mm]
NGC 1808 &    19 &  4.06,  5.78 &  0.42,  0.60  & NGC 4321 &     2 &  4.32,  4.55 &  0.26,  0.28\\*
         &       &  0.18,  0.27 &  0.02,  0.03  &          &       &  0.64,  0.78 &  0.04,  0.05\\[3mm]
NGC 2217 & $-$24 &  3.54,  4.79 &  0.29,  0.39  & NGC 4340 &     4 &  4.24,  5.22 &  0.54,  0.67\\*
         &       &  0.77,  1.08 &  0.06,  0.09  &          &       &  0.38,  0.50 &  0.05,  0.06\\[3mm]
NGC 2642 & $-$30 &  7.01,  7.29 &  0.42,  0.43  & NGC 4503 &    69 &  1.74,  2.05 &  0.22,  0.26\\*
         &       &  0.41,  0.55 &  0.02,  0.03  &          &       &  0.32,  0.53 &  0.04,  0.07\\[3mm]
NGC 2646 &  (74) &  4.33,  5.69 &  0.43,  0.56  & NGC 4725 & $-$84 &  7.18,  7.61 &  0.37,  0.39\\*
         &       &  0.56,  0.89 &  0.06,  0.09  &          &       &  0.45,  0.55 &  0.02,  0.03\\[3mm]
NGC 2681 &       &  4.17,  5.00 &  0.46,  0.55  & NGC 4736 &    60 &  3.27,  4.44 &  0.38,  0.52\\*
         &       &  1.50,  1.58 &  0.17,  0.17  &          &       &  0.34,  0.62 &  0.04,  0.07\\*
         &       &  0.14,  0.28 &  0.02,  0.03  &  & & &\\[3mm]
NGC 2859 &    80 &  4.38,  5.54 &  0.29,  0.37  & NGC 4785 &    81 &  2.80,  3.06 &  0.21,  0.23\\*
         &       &  0.49,  0.75 &  0.03,  0.05  &          &       &  0.36,  0.44 &  0.03,  0.03\\[3mm]
NGC 2950 &    80 &  2.16,  2.79 &  0.37,  0.48  & NGC 4984 &  (25) &  2.94,  3.14 &  0.46,  0.49\\*
         &       &  0.27,  0.33 &  0.05,  0.06  &          &       &  0.36,  0.44 &  0.06,  0.07\\[3mm]
NGC 2962 &    60 &  4.71,  6.98 &  0.41,  0.61  & NGC 5365 & $-$57 &  5.43,  6.74 &  0.41,  0.51\\*
         &       &  0.84,  1.01 &  0.07,  0.09  &          &       &  0.63,  0.90 &  0.05,  0.07\\[3mm]
\hline
\end{tabular}
\medskip

Deprojected relative position angles and sizes for double-barred and
inner-disk galaxies.  For each galaxy, the first line contains values
for the outer bar and the second line contains values for the inner
bar (double-barred galaxies) or inner disk.  Col.\ (1): Galaxy name. 
Col.\ (2): Difference in position angle between inner bar or disk and
outer bar.  Positive = inner bar/disk leads outer bar, negative =
inner bar/disk trails, based on the sense of rotation
(Table~\ref{tab:galaxies-db} or ~\ref{tab:galaxies-id}).  Values in
parentheses are for galaxies where the sense of rotation is undefined. 
No values are given for NGC~2681 because there are, in effect, two
``outer'' bars.  Col.\ (3) Bar/disk linear sizes, in kpc.  The first
number is \amax, the second is \lbar.  Col.\ (4) Bar/disk sizes
relative to galaxy size $R_{25}$; the first number is $\amax/R_{25}$,
the second is $\lbar/R_{25}$.  Note that NGC~3368, 3945, 4785, and
7187 have both inner bars \textit{and} inner disks, and are thus
included in each section.

\end{minipage}
\end{table*}

\setcounter{table}{3}
\begin{table*}[t]
\begin{minipage}{126mm}
\caption{Continued}
\begin{tabular}{lrrr|lrrr}
\hline
Name & $\Delta$PA & Bar/Disk Size & Bar/Disk Size & 
Name & $\Delta$PA & Bar/Disk Size & Bar/Disk Size \\
     & \degr{}    & kpc           & f($R_{25})$   &
     & \degr{}    & kpc           & f($R_{25})$ \\
(1)  & (2)        & (3)           & (4) &
(1)  & (2)        & (3)           & (4) \\
\hline
 & & & & & & &\\
NGC 5728 & $-$47     & 10.91, 13.84 &  0.66,  0.83  & NGC 7716 &    37     &  3.66,  4.16 &  0.35,  0.39\\*
         &           &  0.42,  0.86 &  0.03,  0.05  &          &           &  0.53,  0.92 &  0.05,  0.09\\[3mm]
NGC 5850 & $-$59     & 11.67, 13.89 &  0.53,  0.64  &  Mrk 573 & $-$83     &  2.98,  3.97 &  0.23,  0.30\\*
         &           &  1.14,  1.37 &  0.05,  0.06  &          &           &  0.41,  0.71 &  0.03,  0.05\\[3mm]
NGC 6654 &    77     &  3.71,  5.42 &  0.34,  0.50  &  UGC 524 &    24     &  6.04,  7.82 &  0.28,  0.36\\*
         &           &  0.45,  0.70 &  0.04,  0.06  &          &           &  0.73,  0.80 &  0.03,  0.04\\[3mm]
NGC 6684 &  (60)     &  2.64,  2.84 &  0.33,  0.35  & ESO 215-G031 &     7 &  6.40,  7.92 &  0.56,  0.69\\*
         &           &  0.23,  0.33 &  0.03,  0.04  &          &           &  0.92,  1.07 &  0.08,  0.09\\[3mm]
NGC 6782 & $-$26     &  6.97,  8.36 &  0.45,  0.54  & ESO 320-G030 &    56 &  4.99,  8.02 &  0.38,  0.62\\*
         &           &  1.05,  1.17 &  0.07,  0.08  &          &           &  0.80,  0.92 &  0.06,  0.07\\[3mm]
NGC 7098 & $-$31     &  5.63,  7.85 &  0.34,  0.47  & ESO 443-G039 &  (23) &  2.22,  3.83 &  0.29,  0.50\\*
         &           &  1.24,  1.63 &  0.07,  0.10  &          &           &  0.44,  0.58 &  0.06,  0.08\\[3mm]
NGC 7187 & $-$10     &  3.42,  3.42 &  0.51,  0.51  & ESO 447-G030 & $-$34 &  3.67,  3.95 &  0.47,  0.50\\*
         &           &  0.45,  0.64 &  0.07,  0.09  &          &           &  0.63,  0.81 &  0.08,  0.10\\[3mm]
NGC 7280 &    78     &  1.36,  3.34 &  0.18,  0.43  & IRAS 03565+2139 & 60 &  6.37,  7.35 &  0.77,  0.88\\*
         &           &  0.18,  0.24 &  0.02,  0.03  &          &           &  0.82,  0.97 &  0.10,  0.12\\[5mm]
\hline
\multicolumn{8}{c}{Inner-Disk Galaxies} \\
\hline
 & & & & & & &\\
 NGC 151 & $-$86 &  9.89, 11.53 &  0.38,  0.44  & NGC 4262 &  (46)     &  1.04,  1.27 &  0.25,  0.31\\*
         &       &  1.02,  1.45 &  0.04,  0.05  &          &           &  0.26,  0.33 &  0.06,  0.08\\[3mm]
 NGC 470 & $-$64 &  4.53,  5.61 &  0.36,  0.44  & NGC 4386 &  (10)     &  3.29,  4.74 &  0.34,  0.49\\*
         &       &  0.41,  1.13 &  0.03,  0.09  &          &           &  0.31,  0.42 &  0.03,  0.04\\[3mm]
NGC 1398 &    87 &  4.19,  4.85 &  0.25,  0.29  & NGC 4612 &  (68)     &  1.64,  1.93 &  0.30,  0.35\\*
         &       &  0.23,  0.62 &  0.01,  0.04  &          &           &  0.22,  0.41 &  0.04,  0.07\\[3mm]
NGC 2787 & $-$58 &  1.58,  1.96 &  0.46,  0.57  & NGC 4754 &  (73)     &  3.72,  4.37 &  0.33,  0.39\\*
         &       &  0.66,  0.77 &  0.19,  0.22  &          &           &  0.55,  0.77 &  0.05,  0.07\\[3mm]
NGC 2880 & $-$62 &  1.28,  1.44 &  0.19,  0.22  & NGC 4785 & $-$35     &  2.80,  3.06 &  0.21,  0.23\\*
         &       &  0.32,  0.48 &  0.05,  0.07  &          &           &  1.17,  1.51 &  0.09,  0.11\\[3mm]
NGC 3266 &  (85) &  1.47,  1.92 &  0.26,  0.34  & NGC 7007 &  (72)     &  1.86,  2.28 &  0.18,  0.22\\*
         &       &  0.10,  0.22 &  0.02,  0.04  &          &           &  0.42,  0.53 &  0.04,  0.05\\[3mm]
NGC 3368 & $-$52 &  4.39,  5.40 &  0.38,  0.47  & NGC 7187 &    70     &  3.42,  3.42 &  0.51,  0.51\\*
         &       &  1.09,  1.56 &  0.09,  0.13  &          &           &  1.18,  1.41 &  0.18,  0.21\\[3mm]
NGC 3384 &  (85) &  1.73,  1.96 &  0.19,  0.21  & UGC 6062 &  (64)     &  2.98,  3.72 &  0.48,  0.60\\*
         &       &  0.15,  0.62 &  0.02,  0.07  &          &           &  0.33,  0.49 &  0.05,  0.08\\[3mm]
NGC 3412 &  (66) &  1.18,  1.65 &  0.20,  0.28  & ESO 378-G020 &  (53) &  1.93,  2.32 &  0.27,  0.32\\*
         &       &  0.05,  0.33 &  0.01,  0.06  &          &           &  0.15,  0.24 &  0.02,  0.03\\[3mm]
NGC 3945 & $-$87 &  4.65,  5.67 &  0.32,  0.38  & ESO 443-G017 & $-$67 &  2.36,  2.95 &  0.29,  0.37\\*
         &       &  0.94,  1.68 &  0.06,  0.11  &          &           &  0.44,  0.81 &  0.06,  0.10\\[3mm]
NGC 4143 &    37 &  1.49,  2.45 &  0.28,  0.46  & & & & \\*
         &       &  0.32,  0.48 &  0.06,  0.09  & & & & \\[3mm]
\hline
\end{tabular}
\end{minipage}
\end{table*}

\end{document}